\documentclass[10pt]{article}
\usepackage{amsmath}
\usepackage{amsfonts}
\usepackage{hyperref}
\def\nn{\nonumber}

\def\half{{1\over2}}

\def\xx{{\mathrm{x}}}
\def\yy{{\mathrm{y}}}
\def\zz{{\mathrm{z}}}
\def\kkk{{\mathrm{k}}}

\def\hpsi{{\hat{\Psi}}}
\def\hchi{{\hat{\chi}}}
\def\ha{\hat{a}}
\def\dv{d^{3}}
\def\hnew{\lambda}
\def\csou{c_{sound}}
\def\had{{\hat{a}}^{\dagger}}
\def\phigrav{\Phi_{\mathrm{grav}}}

\def\demi{\frac{1}{2} } 
 \newcommand{\dr}{\rightarrow}   
\newcommand{\R}{\mathbb{R}}   
\newcommand{\mathbold}[1]{\mbox{\boldmath $#1$}}

\newcommand{\Vext}{V_\mathrm{ext}}
\newcommand{\csound}{c_\mathrm{s}}

\newcommand{\im}{\mathrm{i}}

\renewcommand{\v}{\mathbf{v}}

\newcommand{\bnabla}{\mathbold{\nabla}}

\def\lll{{\cal L}} %\def\mm{{\cal M}} \def\nn{{\nonumber}}
\def\im{{\rm i}}
\def\csound{c_{\rm sound}}
\def\Vext{V_{\rm ext}}

 \newcommand{\mone}{^{-1}}
\newcommand{\be}{\begin{equation}}
\newcommand{\bee}{\begin{equation}}
\newcommand{\eeq}{\end{equation}} %\newcommand{\ee}{\end{equation}}

\def\mmm{\mathfrak{m}}
\def\nnn{\mathfrak{n}}
 
\def\ov{\overline}
\newcommand{\ket}[1]{|#1\rangle} \newcommand{\bra}[1]{\langle #1|}

 \newcommand{\bem}{\begin{bmatrix}} \newcommand{\eem}{\end{bmatrix}}
\newcommand{\beq}{\begin{equation}} \newcommand{\ee}{\end{equation}} \newcommand{\eq}{\end{equation}}
\newcommand{\beqa}{\begin{eqnarray}} \newcommand{\eeqa}{\end{eqnarray}}

\def\ie{{i.e. \/}}
\def\eg{{e.g. \/}}
\def\cf{{c.f. \/}}

\def\x{{\bf x}}
\def\mn{{\mu\nu}}
\def\mmm{{\mathfrak{m}}}
\def\hphi{\hat\phi}

\title{Analogue Models for Emergent Gravity\footnote{To appear in the Proceedings of the XVIII  
SIGRAV Conference, Cosenza, September 22-25, 2008}}

\author{S. Liberati$^{1,2}$, F. Girelli$^3$ and L. Sindoni$^{1,2}$\\
\textit{\small {$^1$ SISSA, Via Beirut 2-4, 34151, Trieste (Italy)}}\\
\textit{\small {$^2$ INFN, Sez. di Trieste}}\\
\textit{\small {$^3$ School of Physics, University of Sydney (Australia)}}}

\begin{document}

\maketitle

\begin{abstract}

Gravity stands out among the fundamental interactions because of its apparent incompatibility with having a quantum description. Moreover, thermodynamic aspects of gravitation theory appears as puzzling features of some classical solutions such as black holes. These and other aspects of gravitational theories have recently lead to the proposal that gravity might not be a fundamental interaction but rather an emergent phenomenon, a sort of hydrodynamic limit of some more fundamental theory. In order to further explore this possibility we shall here discuss two systems where such emergence of a gravitational dynamics is observed.  We shall  consider first the case of a non-relativistic Bose-Einstein condensate and then a more abstract model based {on scalar fields living on a Riemannian manifold}. This will allow us to put in evidence the general issues related to emergent gravity scenarios with a particular attention to the role and nature of Lorentz and diffeomorphism invariance.

\end{abstract}

\section{Introduction}

In spite of being the first force of Nature to be understood in physical terms, gravity is somehow still a riddle for  physicists. Not only it keeps evading a full quantum description as well as any form of unification with the other interactions, it also puzzles us with profound questions and unexpected features. We will not attempt here to present a complete list of these startling aspects of gravitation theory, but we can recall, for example, the surprising connection between gravity and thermodynamics associated to black hole physics \cite{Bardeen:1973gs,Hawking:1974sw,Jacobson} as well as the deep questions associated to the nature of inertia and time \cite{Isham,Kuchar,Barbour}.

In recent years, a new approach to these old problems has been gaining momentum and many authors have been advancing the idea that gravity could all in all be an intrinsically classic/large scale phenomenon similar to a condensed matter state made of many atoms~\cite{BLH}. In this sense gravity would not be a fundamental interaction but rather a large scale/number effect, something emergent from a quite different dynamics of some elementary quantum objects. In this sense, many examples can be brought up, starting from the causal set proposal \cite{Bombelli:1987aa}, passing to group field theory \cite{Oriti:2007qd} or the recent quantum graphity models \cite{Konopka:2008hp} and other approaches (see \eg \cite{Dreyer}).

All these models and many others share a common scheme:  they consider a fundamental theory which is not General Relativity and examine, using different techniques often borrowed from condensed matter physics, how space, time and their dynamics could emerge in some regime. It is perhaps important to remark that such ``emergent gravity" scenarios should not be seen as alternatives to quantum gravity proposals (e.g. superstrings theory or loop quantum gravity) rather one should think of them as different incarnations of a more general paradigm about how classical spacetime geometry and dynamics could be recovered from such quantum gravity scenarios.

In this sense a leading inspirational role also been played by a parallel stream of research which goes under the name of ``analogue models of gravity" \cite{LRR}. These are condensed matter systems which have provided toy models showing how at least the concept of a pseudo-Riemannian metric and Lorentz invariance of matter equations of motion can be emergent. For example, non-relativistic systems which admit some hydrodynamics description can be shown to have perturbations (phonons) whose propagation is described, at low energies, by hyperbolic wave equations on an effective Lorentzian geometry~\cite{LRR}. While these models have not provided so far  also an analogy of emergent gravitational dynamics equations they do have provided a new stream of ideas about many other pressing problems in gravitation theory (see for example recent works on the origin of the cosmological constant in emergent gravity~\cite{L-Vol}). The studies presented here should then be considered as exploratory toy models aimed at gaining an understanding of the crucial ingredients necessary for a general emergence paradigm to work.

\section{Analogue models as a test-field for emergent gravity}

Analogue models for gravity have provided a powerful tool for testing (at least in principle) kinematical features of classical and quantum field theories in curved spacetimes \cite{LRR}. The typical setting is the one of sound waves propagating in a perfect fluid \cite{unruh,visseracou}. Under certain conditions, their equation can be put in the form of a Klein-Gordon equation for a massless particle in curved spacetime, whose geometry is specified by the acoustic metric. {Among the various condensed matter systems so far considered, Bose-Einstein condensate (BEC) \cite{Garay:1999sk,Barcelo:2000tg} had in recent years a prominent role for their simplicity as well as for the high degree of sophistication achieved by current experiments. In a BEC system one can consider explicitly the quantum field theory of the quasi-particles (or phonons), the massless excitations over the condensate state, propagating over the condensate as the analogue of a quantum field theory of a scalar field propagating over a curved effective spacetime described by the acoustic metric. It provides therefore a natural framework to explore different aspects of quantum field theory in various interesting curved backgrounds (for example quantum aspects of black hole physics \cite{Balbinot:2004da, BLVsemicls collapse} or the analogue of the creation of cosmological perturbations \cite{Barcelo:2003wu, Silke-Infla}).
Unfortunately, up to now, the analogy with gravity is only partial: there is no analogy with some sort of (semiclassical) Einstein equations, since it has not been possible to put the fluid equations, which are those describing the dynamics of the acoustic metric, in a geometrical form which could eventually lead to a complete dynamical analogy with general relativity \cite{Barcelo:2001tb}.  Our first task here is to show how to fill this gap in the case of BEC and to gather, from the  understanding of the emergence of a BEC analogue gravitational dynamics, general lessons about possible features of ``emergent gravity" scenarios.

\section{Emergent spacetime in BEC: a review}
In BEC, the effective emerging metric depends on the properties of the condensate wave-function. One can expect therefore the gravitational degrees of freedom to be encoded in the variables describing the condensate wave-function \cite{Barcelo:2000tg}, which is solution of the well known Bogoliubov--de Gennes (BdG) equation. The dynamics of gravitational degrees of freedom should then be  inferred from this  equation, which is essentially non-relativistic.  The gravitodynamics of the BEC should therefore be the analogue of some sort of {Newtonian gravity}, and we shall reinterpret the BdG equation as a modified Poisson equation.

The ``emerging matter", the quasi-particles, in the standard BEC, are phonons, \ie~massless excitations. Since we expect the quasi-particles to be the matter source in the Poisson equation, we run a priori into a problem:  massless particles are not treatable in the framework of Newtonian mechanics. To avoid this issue, we shall then introduce a new term in the BEC Hamiltonian  which will softly break the usual $U(1)$ symmetry and therefore will allow the quasi-particles to acquire mass.

\subsection{The hydrodynamic limit}
Let us start by very briefly reviewing the derivation of the acoustic
metric for a BEC system, and show that the equations for the phonons
of the condensate closely mimic the dynamics of a scalar field in a
curved spacetime.  In the dilute gas approximation, one can describe a
Bose gas through a quantum field ${\widehat \Psi}$ satisfying
\begin{eqnarray}
\im\hbar \; \frac{\partial }{\partial t} {\widehat \Psi} =
\left(
  - {\hbar^2 \over 2m} \nabla^2 + \Vext(\x)
  +\kappa(a)\;{\widehat \Psi}^{\dagger}{\widehat \Psi}
\right){\widehat \Psi}. \label{eq:fieldeq0}
\end{eqnarray}
$m$ is the mass of the atoms,  $a$ is the scattering length for the atoms and $\kappa$ parameterises the strength of the interactions between the different bosons in the gas.
It can be re-expressed in terms of the scattering length $a$ as
\begin{equation}
\kappa(a) = \frac{4\pi a \hbar^2}{m}.
\end{equation}
As usual, the quantum field can be separated into
a macroscopic (classical) condensate and a fluctuation: ${\widehat
\Psi}=\psi+{\widehat \varphi}$, with $\langle {\widehat \Psi}
\rangle=\psi $. Then, by adopting the self-consistent mean field
approximation
%(see for example~\cite{Griffin})
%
\begin{eqnarray}
{\widehat \varphi}^{\dagger}{\widehat \varphi}{\widehat \varphi}
\simeq
2\langle {\widehat \varphi}^{\dagger}{\widehat \varphi} \rangle \;
{\widehat \varphi}
+ \langle {\widehat \varphi} {\widehat \varphi}
\rangle \; {\widehat \varphi}^{\dagger},
\end{eqnarray}
one can arrive
at the set of coupled equations:
\begin{eqnarray}
\im \hbar\;\frac{\partial }{\partial t} \psi(t,\x)
&=&
\left ( - {\hbar^2 \over
2m} \nabla^2 + \Vext(\x) + \kappa \; n_c \right) \psi(t,\x)
\nonumber\\
&&
\qquad + \kappa \left\{2\tilde n  \psi(t,\x)+ \tilde m \psi^*(t,\x) \right\};
\label{bec-self-consistent1}
\\
&& \nonumber\\
\im \hbar \; \frac{\partial }{\partial t} {\widehat \varphi}(t,\x)
&=&
\left( - {\hbar^2 \over 2m} \nabla^2  +
\Vext(\x)   +\kappa \;2 n_T \right){\widehat \varphi} (t,\x)
\nonumber
\\
&&
\qquad +  \kappa \; m_T \; {\widehat \varphi}^{\dagger}(t,\x).
\label{quantum-field}
\end{eqnarray}
Here
\begin{eqnarray}
&& n_c \equiv \left| \psi(t,\x) \right|^2;
\quad
m_c \equiv \psi^2(t,\x);
\\
&& \tilde n \equiv  \langle
{\widehat \varphi}^{\dagger}\,{\widehat \varphi} \rangle;
\quad \quad  \quad
\tilde m \equiv  \langle {\widehat \varphi}\, {\widehat \varphi} \rangle;
\\
&& n_T=n_c+\tilde n;
\quad \quad
m_T=m_c+\tilde m.
\end{eqnarray}
In general one will have to solve both equations for $\psi$ and $\widehat \phi$ simultaneously. The equation for the condensate wave function $\psi$ is
closed only when the back-reaction effects due to the fluctuations are neglected. (The back-reaction being hidden in the quantities $\tilde m$ and $\tilde n$.)
This  approximation leads then to the so-called Gross--Pitaevskii equation.

Adopting the Madelung representation for
the wave function $\psi$ of the condensate
\begin{equation}
\psi(t,\x)=\sqrt{n_c(t,\x)} \; \exp[-\im\theta(t,\x)/\hbar],
\end{equation}
and defining an irrotational ``velocity field'' by $\v\equiv{\bnabla\theta}/{m}$, the Gross--Pitaevskii equation can be rewritten
as a continuity equation plus an Euler equation:
\begin{eqnarray}
&& \frac{\partial}{\partial t}n_c+\bnabla\cdot({n_c \v})=0,
\label{E:continuity}\\
&& m\frac{\partial}{\partial t}\v+\bnabla\left(\frac{mv^2}{2}+
V_\mathrm{ext}(t,\x)+\kappa n_c- \frac{\hbar^2}{2m}
\frac{\nabla^{2}\left(\sqrt{n_c}\right)}{\sqrt{n_c}} \right)=0.
\label{E:Euler1}
\end{eqnarray}
These equations are completely equivalent to those of an irrotational
and inviscid fluid apart from the existence of the so-called quantum
potential
\begin{equation}
V_\mathrm{quantum}=
-\hbar^2\nabla^{2}\sqrt{n_c}/(2m\sqrt{n_c}),
\end{equation}
which has the dimensions of an energy. Note that
\begin{equation}
n_c \; \nabla_i
V_\mathrm{quantum} \equiv n_c \; \nabla_i \left[ -{\hbar^2\over2m}
{\nabla^{2}\sqrt{n_c}\over\sqrt{n_c}} \right] = \nabla_j \left[
-{\hbar^2\over4m} \; n_c \; \nabla_i \nabla_j \ln n_c \right],
\end{equation}
which justifies the introduction of the so-called
quantum stress tensor
\begin{equation}
\sigma_{ij}^\mathrm{quantum} =
-{\hbar^2\over4m} \; n_c \; \nabla_i \nabla_j \ln n_c.
\end{equation}
This tensor has the dimensions of pressure, and may be viewed as an
intrinsically quantum anisotropic pressure contributing to the Euler
equation.  If we write the mass density of the Madelung fluid as $\rho
= m \; n_c$, and use the fact that the flow is irrotational then the
Euler equation takes the form
\begin{equation} \rho \left[
\frac{\partial} {\partial t}\v+ (\v\cdot\bnabla) \v \right] + \rho \;
\bnabla \left[\frac{V_\mathrm{ext}(t,\x)}{m} \right] + \bnabla
\left[{\kappa \rho^2\over 2 m^2}\right] + \bnabla \cdot \sigma^\mathrm{quantum}
=0. \label{E:Euler2}
\end{equation}
Note that the term
$V_\mathrm{ext}/m$ has the dimensions of specific enthalpy, while $\kappa
\rho^2/(2m)$ represents a bulk pressure.  When the gradients in the
density of the condensate are small one can neglect the quantum stress
term leading to the standard hydrodynamic approximation. Because the
flow is irrotational, the Euler equation is often more conveniently
written in Hamilton--Jacobi form:
\begin{equation} m
\frac{\partial}{\partial t}\theta+ \left( \frac{[\bnabla\theta]^2}{2m}
+V_\mathrm{ext}(t,\x)+\kappa n_c-
\frac{\hbar^2}{2m}\frac{\nabla^{2}\sqrt{n_c}}{\sqrt{n_c}}
\right)=0. \label{E:HJ} \end{equation}
Apart from the wave function
of the condensate itself, we also have to account for the (typically
small) quantum perturbations of the system
(\ref{quantum-field}).

Let us consider now the quantum perturbations above the condensate. These can be described in
several different ways, here we are interested in the ``quantum
acoustic representation''
\begin{eqnarray}
\widehat
\varphi(t,\x)=e^{-\im \theta/\hbar} \left({1 \over 2
\sqrt{n_c}} \; \widehat n_1 - \im \; {\sqrt{n_c} \over \hbar} \;\widehat
\theta_1\right),
\label{representation-change}
\end{eqnarray}
where
$\widehat n_1,\widehat\theta_1$ are real quantum fields.  By using
this representation Equation~(\ref{quantum-field}) can be rewritten as
\begin{eqnarray}
&&\partial_t \widehat n_1 + {1\over m}
\bnabla\cdot\left( n_1 \; \bnabla \theta + n_c \; \bnabla \widehat
\theta_1 \right) = 0, \label{pt1}
\\ &&\partial_t \widehat \theta_1   +
{1\over m} \bnabla \theta \cdot \bnabla \widehat \theta_1
+ \kappa(a) \; n_1 - {\hbar^2\over2 m}\; D_2 \widehat n_1 = 0.
\label{pt2}
\end{eqnarray}
Here $D_2$ represents a second-order differential operator obtained
from linearizing the quantum potential. Explicitly:
\begin{eqnarray}
D_2\, \widehat n_1 &\equiv&
-\half n_c^{-3/2} \;[\nabla^2
(n_c^{+1/2})]\; \widehat n_1
+\half n_c^{-1/2} \;\nabla^2
(n_c^{-1/2}\; \widehat n_1).
\end{eqnarray}
The equations we have just written can be obtained easily by
linearizing the Gross--Pitaevskii equation around a classical
solution: $n_c \rightarrow n_c + \widehat n_1$, $\phi \rightarrow \phi
+ \widehat \phi_1$.  It is important to realise that in those
equations the back-reaction of the quantum fluctuations on the
background solution has been assumed negligible.  We also see in
Equations~(\ref{pt1}, \ref{pt2}), that time variations of
$V_\mathrm{ext}$ and time variations of the scattering length $a$
appear to act in very different ways.  Whereas the external potential
only influences the background Equation~(\ref{E:HJ}) (and hence the
acoustic metric in the analogue description), the scattering length
directly influences both the perturbation and background equations.
{From} the previous equations for the linearised perturbations it is
possible to derive a wave equation for $\widehat \theta_{1}$ (or
alternatively, for $\widehat n_{1}$). All we need is to substitute in
Equation~(\ref{pt1}) the $\widehat n_{1}$ obtained from
Equation~(\ref{pt2}).  This leads to a PDE that is second-order in
time derivatives but infinite order in space derivatives -- to
simplify things we can construct the symmetric $4 \times 4$ matrix
\begin{equation}
f^{\mu\nu}(t,\x) \equiv
\begin{bmatrix}
   f^{00}&\vdots&f^{0j}\\
   \cdots\cdots&\cdot&\cdots\cdots\cdots\cdots\cr
   f^{i0}&\vdots&f^{ij}\\
\end{bmatrix}.
\label{E:explicit}
\end{equation}
(Greek indices run from
$0$--$3$, while Roman indices run from $1$--$3$.)  Then, introducing
(3+1)-dimensional space-time coordinates
\begin{equation}
x^\mu \equiv (t;\, x^i)
\end{equation}
the wave equation for $\theta_{1}$ is easily rewritten as
\begin{equation}
% \label{E:compact}
\partial_\mu ( f^{\mu\nu} \;
\partial_\nu \widehat \theta_1) = 0. \label{weq-phys}
\end{equation}
Where the $f^{\mu\nu}$ are \emph{differential operators} acting on space
only:
\begin{eqnarray}
f^{00} &=& - \left[ \kappa(a) - {\hbar^2\over 2m}\;
D_2 \right]^{-1} \\ f^{0j} &=& -\left[ \kappa(a) - {\hbar^2\over 2m}\; D_2
\right]^{-1}\; {\nabla^j \theta_0\over m} \\ f^{i0} &=& - {\nabla^{i}
\theta_0\over m} \; \left[ \kappa(a) - {\hbar^2\over2 m}\; D_2
\right]^{-1} \\ f^{ij} &=& {n_c \; \delta^{ij}\over m} - {\nabla^{i}
\theta_0\over m} \; \left[ \kappa(a) - {\hbar^2\over2 m}\; D_2
\right]^{-1}\; {\nabla^{j} \theta_0\over m}.
\end{eqnarray}
Now, if we make a spectral decomposition of the field $\widehat
\theta_1$ we can see that for wavelengths larger than $\xi=\hbar /m\csound$
($\xi$ corresponds to the ``healing length'', as we will explain
below and $\csound(a,n_c)^2={\kappa(a) \; n_c \over m}$), the terms coming from the linearization of the quantum
potential (the $D_2$) can be neglected in the previous expressions, in
which case the $f^{\mu\nu}$ can be approximated by scalars, instead of
differential operators. (This is the heart of the acoustic
approximation.)  Then, by identifying
\begin{equation}
\sqrt{-g} \; g^{\mu\nu}=f^{\mu\nu},
\end{equation}
the equation for the field $\widehat \theta_1$
becomes that of a (massless minimally coupled) quantum scalar field
over a curved background
\begin{equation}
\Delta\theta_{1}\equiv\frac{1}{\sqrt{-g}}\;
\partial_{\mu}\left(\sqrt{-g}\; g^{\mu\nu}\; \partial_{\nu}\right)
\widehat\theta_{1}=0,
\end{equation}
with an effective metric of the form
\begin{equation} g_{\mu\nu}(t,\x) \equiv {n_c\over m\;
\csound(a,n_c)}
\begin{bmatrix}
   -\{\csound(a,n_c)^2-v^2\}&\vdots& - v_j \\
   \cdots\cdots\cdots\cdots&\cdot&\cdots\cdots\\
   -v_i&\vdots&\delta_{ij}\\
\end{bmatrix}.
\end{equation}
Here the magnitude
$\csound(n_c,a)$ represents the speed of the phonons in the medium:
\begin{equation}
\csound(a,n_c)^2={\kappa(a) \; n_c \over m},
\end{equation}
{and $v_i$ is the velocity field of the fluid flow,}
\begin{equation}
v_i= \frac{1}{m}\nabla_i \theta.
\end{equation}

\subsection{Lorentz violation in BEC}

It is interesting to consider the case in which the above
``hydrodynamical" approximation for BECs does not hold. In order to
explore a regime where the contribution of the quantum potential
cannot be neglected we can use the so called {\emph{eikonal}}
approximation, a high-momentum approximation where the phase
fluctuation $\widehat \theta_1$ is itself treated as a slowly-varying
amplitude times a rapidly varying phase. This phase will be taken to
be the same for both $\widehat n_1$ and $\widehat \theta_1$
fluctuations. In fact, if one discards the unphysical possibility that
the respective phases differ by a time varying quantity, {any
time-independent difference can be safely reabsorbed in the definition of
the (complex) amplitudes $\mathcal{A}_\theta,\, \mathcal{A}_\rho$.}  Specifically, we shall write
\begin{eqnarray}
{\widehat\theta}_1(t,\x)
&=& \mathrm{Re}\left\{\mathcal{A}_\theta \; \exp(-i\phi) \right\},\\
{\widehat n}_1(t,\x)
&=& \mathrm{Re}\left\{\mathcal{A}_\rho \; \exp(-i\phi)\right\}.
\end{eqnarray}
As a consequence of our starting assumptions, gradients of the
amplitude, and gradients of the background fields, are systematically
ignored relative to gradients of $\phi$.  (Warning:  What we are doing
here is not quite a ``standard'' eikonal approximation, in the sense
that it is not applied directly on the fluctuations of the field
$\psi(t,\x)$ but separately on their amplitudes and phases $\rho_{1}$
and $\phi_{1}$.)  We adopt the notation
\begin{equation}
\omega ={\partial\phi\over\partial t};\qquad k_i = \nabla_i \phi.
\end{equation}
Then the operator $D_2$ can be approximated as
\begin{eqnarray}
D_2 \;{\widehat n}_1
&\equiv& -\half n_c^{-3/2} \;[\Delta (n_c^{+1/2})]\; {\widehat n}_1
+\half n_c^{-1/2} \;\Delta (n_c^{-1/2} {\widehat n}_1)
\\
&\approx&
+\half n_c^{-1} \;[\Delta {\widehat n}_1]
\\
&=& -\half n_c^{-1}
\;k^2 \;{\widehat n}_1.
\end{eqnarray}
A similar result holds for
$D_2$ acting on ${\widehat \theta}_1$.  That is, under the eikonal
approximation we effectively replace the {\emph{operator}} $D_2$ by
the {\emph{function}}
\begin{equation}
D_2 \to -\half n_c^{-1}k^2.
\end{equation}
For the matrix $f^{\mu\nu}$ this effectively
results in the replacement
\begin{eqnarray}
f^{00}
&\to& - \left[\kappa(a) + {\hbar^2 \; k^2\over4m\;n_c} \right]^{-1} \\
f^{0j}
&\to& -\left[ \kappa(a) + {\hbar^2 \; k^2\over4m\;n_c}\right]^{-1}\;
{\nabla^j \theta_0\over m} \\
f^{i0}
&\to& - {\nabla^i \theta_0\over m} \;
\left[ \kappa(a) + {\hbar^2 \; k^2\over4m\;n_c} \right]^{-1} \\
f^{ij}
&\to& {n_c \; \delta^{ij}\over m} - {\nabla^i \theta_0\over m}\;
\left[ \kappa(a) + {\hbar^2 \; k^2\over4m\;n_c} \right]^{-1}\;
{\nabla^j \theta_0\over m}\,.
\end{eqnarray}
(As desired, this has the net effect of making $f^{\mu\nu}$ a matrix
of numbers, not operators.) The physical wave equation~(\ref{weq-phys})
now becomes a nonlinear dispersion relation
\begin{equation}
f^{00} \;\omega^2 + (f^{0i} +f^{i0}) \;\omega \;k_i +
f^{ij} \;k_i \;k_j = 0.
\end{equation}
After substituting the approximate $D_2$ into this dispersion relation
and rearranging, we see (remember: $k^2 = ||k||^2 = \delta^{ij}
\;k_i \;k_j$)
\begin{equation}
-\omega^2 + 2 \; v_0^i \; \omega k_i
+ {n_c k^2\over m}\left[\kappa(a)+{\hbar^2\over4m n_c} k^2\right] -
(v_0^i \; k_i)^2 = 0.
\end{equation}
That is (with $v_0^i=\frac{1}{m}\nabla_i \theta_0$)
\begin{equation}
\left(\omega - v_0^i \; k_i\right)^2 = {n_c k^2\over
m}\left[\kappa(a)+{\hbar^2\over4m n_c} k^2\right]\,.
\end{equation}
Introducing the speed of sound $\csound$ this takes the form:
\begin{equation} \omega= v_0^i \; k_i \pm \sqrt{\csound^2 k^2+\left({\hbar
\over 2 m}\;k^2\right)^2}.
\label{eq:disprel}
\end{equation}

Having described the effective metric in BEC and its limit of applicability we are now in hand, the analogy is fully
established, and one is now in a position to start asking more specific
physics questions.

\section{Emergent gravity in BEC}

The program of extracting some sort of Poisson equation out of the Bogoliubov--de Gennes formalism cannot be carried on in a standard BEC.
Indeed, phonons are massless excitations, and hence, since we want them to enter the Poisson equation as a source term, we have to circumvent the impossibility of treating massless particles in the framework of Newtonian mechanics. This is easily done by making phonons massive. Concretely, this is done by introducing a new term in the Hamiltonian  which will softly break the usual $U(1)$ symmetry associated to number conservation and therefore will allow the quasi-particles to acquire a mass. Essentially, the quasi-particles will be pseudo-Goldstone bosons \cite{BurgessGoldstone,WeinbergQFT1}: their spectrum, instead of being gapless, is gapped. One can expect, then, that instead of massless quasi-particles, the collective modes above the condensate will be massive.

In order to do so, the standard Hamiltonian $\hat{H}_0$ described previously needs to be slightly modified,
by introducing a term which is (softly) breaking the $U(1)$ symmetry in
\eqref{eq:fieldeq0}.
\beq \hat H_0 \dr  \hat H = \hat H_0+ \hat H_\lambda, \qquad \hat H_\lambda = - \frac{\lambda}{2} \int \dv\xx \left( \hpsi(\xx) \hpsi(\xx) + \hpsi^{\dagger}(\xx) \hpsi^{\dagger}(\xx) \right).\label{sy-br} \ee
The parameter $\lambda$  has the same dimension as $\mu$. With this new Hamiltonian, the non-linear equation \eqref{eq:fieldeq0} becomes
\begin{equation}\label{eq:fieldeq}
i\hbar \frac{\partial }{\partial t}\hpsi = [\hat H , \hpsi] =  -\frac{\hbar^2}{2m} \nabla^{2}\hpsi - \mu \hpsi  + \kappa |\hpsi|^{2}\hpsi- \lambda \hpsi^{\dagger}.
\end{equation}

The addition of this term implies that the whole dynamics of the system must be reconsidered.
The analysis presented in the subsection \eqref{sec:quasipart} will show how $\hat H_\lambda$ generates  a mass for the quasi-particle. Even though $\hat H_\lambda$  both creates and destroys pairs of atoms, it is not difficult to check that $\hat H_\lambda$ is not commuting with the number operator $\hat N$,
\begin{equation}\label{eq:commutatorHN}
[\hat{H}_{\lambda},\hat{N}] =
-\lambda \int \dv \xx \left( \hpsi(\xx)\hpsi({\xx}) -\hpsi^{\dagger}(\xx)\hpsi^{\dagger}(\xx)\right)
\end{equation} while unitarity is preserved. In fact, when  applied on a state with a definite number of atoms $n$ we have:
\begin{equation}\label{eq:number}
\ket{n} \rightarrow \ket{n-2} + \ket{n+2},
\end{equation}
which means that an eigenstate of the number operator evolves into a superposition of states with different occupation numbers.
However, the expectation value of the number of operator on its eigenstates is still constant
\begin{equation}
i \hbar \frac{\partial}{\partial t} \bra{n}\hat{N} \ket{n} =  \bra{n}[\hat{N},\hat{H}]\ket{n} = \bra{n}[\hat{N},\hat{H}_{\hnew}]\ket{n}  \propto \langle n | n-2 \rangle -\langle n | n+2 \rangle =0.
\label{eq:Nav}
\end{equation}

Given this crucial difference with the standard description of BECs, a formalism like the particle-number-conserving one \cite{Gardiner,castindum} cannot be immediately used for these new models. However, given the important improvements in the description of inhomogeneous condensates provided by this formalism it will be interesting to extend the particle-number-conserving method in a suitable way, in order to be able to control to which accuracy we can trust the standard mean field approximation we are using. In fact, we are assuming that the addition of the new term in the Hamiltonian will not destroy the stability properties of the mean field theory, and in particular the fact that the mean field theory, for nearly homogeneous condensates and trapping potentials, in weakly time-dependent regimes, does offer a good description of the condensate dynamics.

{This point of view (and its implications for the condensation mechanism) deserves some further specifications. While the condensation process can be easily understood as a macroscopic occupation number of an energy level, there are several approaches to describe it mathematically. The mean field approach is particularly convenient: we say that the system of $N$ bosons has condensed whenever the field $\hpsi$ develops a non-zero vacuum expectation value (vev)
\begin{equation}
\bra{\Omega} \hpsi \ket{\Omega} = \psi,
\end{equation}
where $\psi$ is the condensate wave-function. If this mean field is non-vanishing, we have that the two point correlation function
\begin{equation}\label{longrangeorder}
 G(\xx,\yy) = \langle \hpsi^{\dagger}(\xx) \hpsi(\yy) \rangle \approx \psi^*(\xx)\psi(\yy),
\end{equation}
tends to a non-zero constant when $\xx,\yy$ are infinitely separated, \ie the system develops long range correlations \cite{blackbook}.

The mean field method is based on the assumption that the ground state of the system is not the vacuum state of the Fock space, $|0\rangle$, but rather it is similar to a coherent state.
% \cite{Klauder}.
For a single mode Fock space, a coherent state is defined to be:
\begin{equation}
|z\rangle = e^{-|z|^{2}/2} e^{-z \hat{b}^{\dagger}} |0\rangle,
\end{equation}
and it is easy to see that it is an eigenstate of the annihilation operator:
\begin{equation}
\hat{b}|z\rangle = z |z\rangle.
\end{equation}
In the case of BEC, the fact that the state $u_{0}(\xx)$ is macroscopically occupied (\ie there are $N_{0}$ bosons in the state $0$, with $N_{0}/N\approx 1$) can be formalized by taking:
\begin{equation}
|\Omega \rangle \approx e^{-N_{0}/2} e^{-i(N_{0})^{1/2} \hat{a}^{\dagger}_{0}} |0\rangle.
\end{equation}
On this states, the field operator $\hpsi$ behaves like a c-number:
\begin{equation}
 \hpsi(\xx) |\Omega \rangle \approx \sqrt{N_0} u_0(\xx) |\Omega \rangle,
\end{equation}
where the approximation is due to the fact that interactions are introducing some corrections.
This property of the ground state motivates the splitting of the field operators into the classical part, which deals with the condensed phase, and small residual fluctuations, describing the states which are close to the ground state:
\begin{equation} \label{meanfield}
\hpsi = \psi \mathbb{I} +\hchi.
\end{equation}
This is the basic idea of the mean field approximation: the field operator $\hpsi$ is approximated by a classical field, which is describing the condensate, while the fluctuations, assumed to be small, are still encoded in a field operator.

The fact that the solutions to the Gross--Pitaevski equation leads to a non-vanishing mean field, in light of the discussion about the realization of a regime of long range correlations (see \eqref{longrangeorder}), ensures that a condensation has taken place. In this sense, the addition of the new term into the Hamiltonian should not forbid the condensation, provided that $\lambda,\mu,\kappa$ are such that the condensate wavefunction can be different from zero.
}

From a different point of view, one can imagine to keep $\lambda$ very small with respect to all the other energy scales present in the theory, making the new term a tiny perturbation of the system. Of course, nonperturbative effects can spoil this picture. 

The $U(1)$ symmetry in standard BEC is related to the fact that the number of atoms is conserved. The breaking of this symmetry therefore is connected with the failure of this charge to be conserved. There are at least two possible physical implementations of this.

A rather natural option is to have an open system. Concretely, one could imagine to have a condensate which is able to exchange particles with some sort of reservoir, in such a way to preserve, on average, their number. Several settings in this sense could be conceived, {\em e.g.} with coupling with suitably tuned lasers.

A second important case is the one in which the constituents themselves are some sort of collective degrees of freedom. This could be the case, for instance, of the excitations in the so-called quantum Heisenberg ferromagnet defined with a spin system. In this case, the fundamental operators give rise to effective degrees of freedom, called magnons, whose Hamiltonian, in general, is not $U(1)$ invariant. For more details see \cite{magnon}, and references therein.

\subsection{The condensate wave-function}\label{sec:cond}
We consider the dynamics generated by \eqref{eq:fieldeq}, from which we want {to} extract the equation of motion  for the condensate $\psi$.  The evolution of the mean field $\psi$ is easily determined in terms of the eigenstates $\ket{E}$ of the Hamiltonian $\hat H$:
\begin{equation*}
i\hbar \frac{\partial}{\partial t} \psi =i\hbar \frac{\partial}{\partial t } (\bra{E} \hpsi \ket{E})
=\bra{E} i\hbar \frac{\partial}{\partial t} \hpsi \ket{E} =\end{equation*}
\begin{equation} =-\frac{\hbar^{2}}{2m } \nabla^{2} \psi - \mu \psi -\lambda \psi^{*} + \kappa |\psi|^{2} \psi + 2 \kappa \nnn_{E} \psi + \kappa \mmm_{E} \psi^{*},
\label{BdG}
\end{equation}
where $\mmm_{E}=\bra{E}\hchi^{2}\ket{E}, \nnn_{E}=\bra{E}\hchi^{\dagger}\hchi\ket{E}$ encode the effect of the non-condensate atoms.
This is the generalization of the Bogoliubov-de Gennes (BdG) equation for the condensate wave-function to the case $\lambda\neq 0$.

In the standard case, if we have $N$ particles in the condensate, the number density of the non-condensate fraction can be estimated to be of order $1/N$ with respect to the number density of the condensate, when the condensation mechanism is particularly effective. In this sense, the terms $\mmm,\nnn$ are of order $1/N$. This can be safely exported to our case, with the slight modification of the meaning of the number $N$, which does represent only the average number of particles in the condensate (see \eqref{eq:Nav}).

At zeroth order in the $1/N$ expansion, we have   the generalization of the Gross--Pitaevski (GP) equation:
\begin{equation}\label{eq:GP}
i\hbar \frac{\partial}{\partial t} \psi = -\frac{\hbar^{2}}{2m } \nabla^{2} \psi - \mu \psi -\lambda \psi^{*} + \kappa |\psi|^{2} \psi.
\end{equation}
The time independent homogeneous solution to the GP equation is
\begin{equation}\label{eq:condensatedens}
n_{c} = |\psi|^{2} = \frac{\mu+\lambda}{\kappa},
\end{equation}
where we have fixed the phase of the condensate to be zero. In section \ref{subs:fluid} we will show that this is not an arbitrary choice, but rather a consequence of the situation we want to describe once we impose a stability condition for the quasi-particles.

As in the standard case, we define the healing length $\xi$ as the length scale at which the kinetic term is of the same order of magnitude of the self-interaction term in the Hamiltonian:
\begin{equation}\label{healing}
\frac{\hbar^{2}}{2m \xi^{2}} = \kappa n_{c} \Leftrightarrow \xi^{2} = \frac{\hbar^{2}}{2m \kappa n_{c}}.
\end{equation}
Again,
this length represents the spatial scale needed for the condensate to pass from the value $n_c=0$ at the boundary of the region where it is confined to the bulk value $n_c$ (see \cite{Fetter}). Also in this case, the healing length represents the scale of the dynamical processes involving the deformation of the condensate wavefunction. This will have a crucial impact on the emergent gravitational dynamics in these systems.

\subsection{Quasi-particles}\label{sec:quasipart}
The equation of motion for the particles out of the condensate is obtained by subtracting the equation for the condensate \eqref{eq:GP} from the equation for $\hpsi$  given in \eqref{eq:fieldeq}.
We are interested in the propagating modes, so we neglect the self-interactions. We obtain:
\begin{equation}\label{eq:qpdynamics}
i\hbar\frac{\partial}{\partial t} \hchi = -\frac{\hbar^{2}}{2m }\nabla^{2} \hchi + (2 \kappa |\psi|^{2}-\mu) \hchi + (\kappa \psi^{2}-\lambda) \hchi^{\dagger}.
\end{equation}
Let us consider the case of homogeneous condensate with density $n_{c}$ given above.
In this situation we have:
\begin{equation}
i\hbar\frac{\partial}{\partial t} \hchi = -\frac{\hbar^{2}}{2m }\nabla^{2} \hchi + (\mu+2\lambda) \hchi + \mu \hchi^{\dagger}.
\end{equation}
If we decompose the field $\hchi$ in its plane wave components, we can rewrite this equation as
\begin{equation}
i\hbar\frac{\partial}{\partial t} \ha_{\kkk} = \frac{\hbar^{2} k^{2}}{2m} \ha_{\kkk} + (\mu+2\lambda)\ha_{\kkk} + \mu \had_{-\kkk}.
\end{equation}
The mixing between $\ha$ and $\had$ due to the evolution in time becomes then apparent. We therefore pass to the quasi-particle operators $\hphi(\xx)$
\begin{equation}
\hphi(\xx) = \frac{1}{\sqrt{V}} \sum_{\kkk} \hat{b}_{\kkk} e^{i\kkk\cdot \xx},
\end{equation}
which are related to the particle operators through the Bogoliubov transformation
\begin{equation}\label{bogolobov transf}
 \ha_{\kkk} = \alpha(k) \hat{b}_{\kkk} + \beta(k)\hat{b}_{-\kkk}, \quad \textrm{with } \alpha^2(k)-\beta^2(k)=1.
\end{equation}
The coefficients $\alpha,\beta$ are only functions of $k=|\vec\kkk|$, since the condensate is homogeneous and isotropic. The equation of evolution for the quasi-particles is then given by
\begin{equation}
 i\hbar \frac{\partial }{\partial t} \hat{b}_{\kkk} = \mathcal{E}(k)\hat{b}_{\kkk},
\end{equation}
with the energy
\begin{equation}\label{eq:qparticledr}
\mathcal{E}(k) = \left( \frac{\hbar^{4} k^{4}}{4m^{2}} + 4\lambda (\mu+\lambda) + \frac{\mu+2\lambda}{m} \hbar^{2}k^{2}  \right)^{1/2}.
\end{equation}
The Bogoliubov coefficients are given by:
\begin{equation}\label{Bogoliubovcoeff}
\alpha^{2}(k) = \frac{A(k)+\mathcal{E}(k)}{2 \mathcal{E}(k)},
 \qquad \beta^{2}(k) = \frac{1}{2\mathcal{E}(k)}\frac{\mu^2}{A(k)+\mathcal{E}(k)},
\end{equation}
where we have introduced the quantity
\begin{equation}
 A(k) = \frac{\hbar^{2}k^2}{2m} + \mu + 2\lambda.
\end{equation}
The high energy limit of these coefficients is:
\begin{equation}
\lim_{k \rightarrow \infty} \alpha^{2}(k) = 1, \qquad \lim_{k\rightarrow \infty} \beta^{2}(k)= 0,
\end{equation}
which means that  at large wave-number (and hence large momentum), the quasi-particle operators coincide with the particle operators. This matches the behavior of the energy, which becomes just the energy of a non-relativistic particle of mass $m$, just like a free atom.
The dispersion relation \eqref{eq:qparticledr} suggests the introduction of the following quantities:
\begin{equation} \label{eq:soundspeedmass}
\csou^{2} =\frac{\mu + 2 \lambda}{m},  \qquad \mathcal{M}^{2} = 4\frac{\lambda(\mu+\lambda)}{(\mu+2\lambda)^{2}}m^{2}.
\end{equation}
Here $\csou$ plays the role of the speed of sound, while $\mathcal{M}$ plays the role of a rest mass for the quasi-particle. Since ${\mathcal{M}}$ is proportional to $\lambda$, we clearly see that it is the  term $\hat H_\lambda$ that generates the mass of the quasi-particle. When  $\lambda\dr0$, that is when $\hat H\dr \hat H_0$, the quasi-particle becomes massless, \ie a phonon, and the speed of sound reduces to the usual one in BEC. Perturbation theory, therefore, should be a viable strategy to compute the various physical properties of these systems.

Notice that, in order to have a non-negative mass square term, and to avoid a tachyonic instability, we have to require $\lambda\geq0$. In standard BEC, one usually assumes that the chemical potential $\mu$ is positive: indeed if it were negative, there could not be any condensation. In our case, we can relax this requirement and obtain that $\mu>-\lambda$ as a condition. In the following we consider $\mu>0$, in order to be able to consider the case in which the correction we are inserting is very small, without affecting dramatically the condensation. {Indeed, it is easy to see that a condensation can take place even in a system with this soft $U(1)$ breaking by checking the behavior of the two points correlation function $G(\xx,\yy)=\langle \hpsi^{\dagger}(\xx) \hpsi(\yy)\rangle$. It is immediate to realize that, in the case of homogeneous condensate, this correlation function describes long range correlations, since the mean field $\psi$ is non-vanishing (cf. equation \eqref{eq:condensatedens}).}

${\mathcal{M}}$ is proportional to $m$, the mass of the atoms. Defining the ratio $\zeta= \lambda/\mu$, we  introduce the function $F(\zeta)$
\begin{equation}
{\mathcal{M}}^2= F(\zeta)m^2=4 \frac{\zeta(1+\zeta)}{(1+2\zeta)^{2}}m^2.
\end{equation}
Under our assumptions, we have that $\zeta \geq 0$. It is then straightforward to check that on this domain $F(\zeta)$ is a monotonic (increasing) function and that
\begin{equation}
F(0)=0, \qquad \lim_{\zeta\rightarrow + \infty}F(\zeta) = 1.
\end{equation}
We conclude therefore  that the mass of the quasi-particles ${\mathcal{M}}$ is always bounded by the mass of the atoms, ${\mathcal{M}}\in [0,m)$.

It is also interesting to notice that using the variable $\zeta$, the speed of sound is:
\begin{equation}
\csou^{2} = \frac{1 + 2\zeta}{1+\zeta} \frac{\kappa n_{c}}{m}.
\end{equation}
For $\zeta$ small, we then have $\csou^{2} \approx \kappa n_{c}/m$, which is the standard result, while, for $\zeta\rightarrow \infty$, $\csou^{2}\rightarrow 2 \kappa n_{c}/m$.

\subsection{The various regimes for the modified dispersion relation} \label{sec:regimes}
Before moving on to the gravitational dynamics, let us  discuss briefly the content of the dispersion relation  \eqref{eq:qparticledr} for the quasi-particles, rewritten using $\csou$ and ${\mathcal{M}}$.
\begin{equation}\label{eq:qparticledr2}
\mathcal{E}(p) = \left(\frac{p^{4}}{4m^{2}} + \csou^{2} p^{2} + {\mathcal{M}}^{2}\csou^{4}\right)^{1/2},
\end{equation}
where we are using the obvious notation $p=\hbar k$ to simplify the shape of the expressions.
Let us define the  characteristic momenta $p_A$, $p_B$ and $p_C$ such that
\begin{equation}
\frac{p_{A}^{4}}{4m^{2}} = \csou^{2} p_{A}^{2}, \qquad \frac{p_{B}^{4}}{4m^{2}} = {\mathcal{M}}^{2} \csou^{4},
\qquad \csou^{2} p_{C}^{2} = {\mathcal{M}}^{2} \csou^{4},
\end{equation}
so that they are explicitly
\begin{equation}
p_{A}^{2} = 4 m^{2} \csou^{2}, \qquad p_{B}^{2} = 2 m {\mathcal{M}} \csou^{2},
\qquad p_{C}^2={\mathcal{M}}^{2} \csou^{2}.
\end{equation}
They are related through the relations
\begin{equation}
p_{C}^{2} = 2 F(\zeta) p_{B} ^{2}=
4 F^{2}(\zeta) p_{A}^{2}.
\end{equation}
If $\zeta \ll 1$, which will be the regime we shall consider,   we have also that
\begin{equation}\label{regimes}
p_{C}\ll p_B  \ll p_{A}.
\end{equation}
Taking into account \eqref{regimes}, the  characteristic momenta define different regimes:
\begin{itemize}
\item If $p \gg p_A$, the  term $p^4$ dominates, the dispersion relation \eqref{eq:qparticledr2} is well approximated by $\mathcal{E} \sim p^2/2m$, we are in the trans-phononic regime.

\item If on the contrary we have $p_C \ll p\ll p_A$, we can safely neglect the term of order $p^4$, we are then in the relativistic regime since the dispersion relation \eqref{eq:qparticledr2} is well approximated by $\mathcal{E} \sim (p^2\csou^2 + {\mathcal{M}}^2\csou^4)^\demi$. The quasi-particle is then relativistic, when the speed of sound $c_{s}$ is playing the role of the speed of light.
\item If we are in the regime where $p\ll p_{C}$, this means that the quasi-particle has a speed much smaller than $\csou$, so that this is the Galilean limit of the relativistic regime. We are then dealing with a Galilean quasi-particle. The rest mass ${\mathcal{M}} \csou^2$ provides the usual constant shift of the Galilean energy $\mathcal{E}\sim {\mathcal{M}} \csou^2 +  p^2/{2\mathcal{M}}  $.
\end{itemize}

\subsection{The fluid description}\label{subs:fluid}
We have already seen that the standard Gross--Pitaevski (GP) equation describing a BEC admits an interesting fluid interpretation, through the Madelung representation.
We are considering now the GP equation given in \eqref{eq:GP}
\begin{equation}
i\hbar \frac{\partial}{\partial t} \psi = -\frac{\hbar^{2}}{2m } \nabla^{2} \psi - \mu \psi -\lambda \psi^{*} + \kappa |\psi|^{2} \psi,
\end{equation}
and we want to use the Madelung representation for the complex field $\psi$:
\begin{equation}
\psi = \sqrt{n_{c}} e^{-i\theta/\hbar}.
\end{equation}
When replacing this into the GP equation, dividing by the phase and splitting the resulting expression into the real and imaginary parts we obtain two equations:
\begin{equation}
\dot{n}_{c} + \vec{\nabla} \cdot (n_{c} \vec{v}) = - \frac{\lambda}{\hbar} n_{c} \sin \left( \frac{2\theta}{\hbar} \right),
\end{equation}
\begin{equation}
\dot{\theta} = V_{\mathrm{quantum}} + \frac{m}{2}v^{2} - \mu-\lambda \cos\left( \frac{2\theta}{\hbar} \right) - \kappa n_{c},
\end{equation}
where we have introduced the velocity field $\vec{v}= -\vec{\nabla}\theta/\hbar$, and
\begin{equation}
 V_{\mathrm{quantum}}  = -\frac{1}{\sqrt{n_{c}}} \frac{\hbar^{2}}{2m} \nabla^{2} \sqrt{n_{c}},
\end{equation}
is the familiar quantum potential term.
These two equations, in the case $\lambda=0$, reduce to the usual form of the continuity equation and the Euler equation for a perfect fluid, once we neglect the quantum potential term.  On the other hand, when $\lambda\neq 0$ the $U(1)$ invariance is broken, and the number operator is no more conserved by the Hamiltonian evolution.

It is interesting to see what happens when we consider the case of homogeneous condensates, $\partial_{\mu} n_{c}=\partial_{\mu} v^{i}=0$. From the first equation we get:
\begin{equation}
\sin \left( \frac{2\theta}{\hbar} \right) = 0 \Leftrightarrow \theta= \frac{l \pi}{2} \hbar, \,\,\,\,\, l\in \mathbb{Z}.
\end{equation}
This result implies that not only $\vec{v}$ is constant, but it actually vanishes. Inserting this result in the second equation we obtain:
\begin{equation}
n_{c} = \frac{\mu + \cos(l\pi) \lambda}{\kappa}.
\end{equation}
The analysis of the quasi-particle dynamics in the case of homogeneous condensates has shown that the case $\lambda<0$ corresponds to a negative mass square term, \ie tachyonic behavior: the energy of a quasi-particle would get an imaginary part leading to exponential growing and damping of modes. Since a choice of the phase of the condensate such as $\cos(l \pi)=-1$ would be completely equivalent to a change of sign of $\lambda$, thus leading to instabilities,
{it is clear that $\cos(l\pi)= 1$ is required for the stability of the condensate}.

\subsection{Gravitational dynamics}\label{sec:grav}
The next step is the analysis of the inhomogeneous condensate, and hence the promised emergence of a gravitational dynamics.
To simplify further the treatment it is better to consider the case of condensates which are nearly, but not exactly, homogeneous:  this will correspond to the case of weak gravitational field. This limitation is consistent with the formalism we are using. Indeed the mean field method certainly is not a good approximation in regions where there are large variations in density (see for instance vortex cores).

In an asymptotically flat spacetime, in order to identify the Newtonian gravitational potential it is necessary to evaluate the non-relativistic limit of the geodesic equation in a weak gravitational field  \cite{MTW}. In the asymptotic region there is a coordinate system such that the metric can be written as $g_{\mu\nu}=\eta_{\mu\nu}+h_{\mu\nu}$. The tensor $h_{\mu\nu}$ encodes the deviation from exact flatness, \ie the gravitational field. In this regime, it is easy to show that the Newtonian gravitational field is identified with the component $h_{00}$:
\begin{equation}
\Phi_{N}(\xx) = -\frac{1}{2} h_{00}(\xx).
\end{equation}
In the context of standard BEC (\ie dealing with the non-linear equation \eqref{eq:fieldeq0}), the quasi-particles travel in an emergent metric $ds^2$ determined in terms of the homogenous condensate $\psi$.
\begin{equation}\label{eq:acousticlineelementasym}
ds^{2} =\frac{n_{c}}{m c_{s}} \left[-\left(c^{2}_{s}- v^{2} \right) dt^{2} -2 v_{i} dt d\xx^{i} +\delta_{ij} d\xx^{i}d\xx^{j} \right],
\end{equation}
where $m$ is the mass of the atoms and $c_s$ and $\vec v$  depend on the properties of the condensate $\psi=\sqrt{n_c}e^{i\theta}$, through $$c_{s}=\frac{\kappa n_c}{m}, \qquad \vec v=\frac{1}{m}\vec \nabla \theta.$$
Considering that the condensate is homogenous, the density and velocity   profiles become constant, \ie respectively   $n_c=n_\infty$, $\vec v=\vec v_{\infty}$.
With the coordinate transformation,
\begin{equation} \label{eq:pgtominkowski}
dT=dt, \qquad dX^{i} = d\xx^{i} - v^{i}_{\infty}dt,
\end{equation}
the line element \eqref{eq:acousticlineelementasym} is rewritten as:
\begin{equation}
ds^{2}_{\infty} = -c_{\infty}^{2} dT^{2} + d\bf{X}^{2}.
\end{equation}

The condition of asymptotic flatness for the acoustic spacetime can be translated into the condition of asymptotic homogeneity for the condensate.  We require then that only in a small region of space, in the bulk, the condensate deviates from perfect homogeneity.

We consider therefore some small deviation from the asymptotic values of the velocity and of the density:
\begin{equation}\label{eq:expansion}
n_{c}=n_{\infty}(1 + 2 u(\xx)), \qquad \vec v = \vec v _{\infty} + \vec w(\xx), \qquad  \textrm{with } u\ll 1, \quad  w\ll v.
\end{equation}
 This implies in particular a rescaling of the speed of sound.
$$ c_{s}^{2} = \frac{\kappa n_{c}}{m} = c_{\infty}^{2} (1 + 2u(\xx)).$$
The acoustic line element \eqref{eq:acousticlineelementasym} becomes then
\begin{equation}
ds^{2}= \frac{n_{c}}{m c_{s}} \frac{m c_{\infty}}{n_{\infty}} \left( -(c_{s}^{2}-v^{2})dt^{2} - 2 v_{i} d\xx^{i}dt + \delta_{ij}d\xx^{i}d\xx^{j} \right),
\end{equation}
where we have introduced a constant prefactor $m c_\infty /n_{\infty}$ in order to have   the conformal factor  asymptotically normalized to one.
Using \eqref{eq:expansion}, together with the coordinate change \eqref{eq:pgtominkowski},  the acoustic line element has the form:
\begin{equation}
ds^{2} = ds^{2}_{\infty} - 3 u(X) c_{\infty}^{2} dT^{2} - 2 w_{i}(X)dT dX^{i} + u(X)\delta_{ij} d X^{i}dX^{j},
\end{equation}
at first order in $u, w_{i}$. Consequently, we see that
\begin{equation}
h_{00}(X) = -3 c_{\infty}^{2}u(X),
\end{equation}
so that  the gravitational field is encoded in the \emph{number density perturbation} of the condensate wave-function $\psi$,
\begin{equation}
\Phi_{N}(X) = \frac{3}{2} c_{\infty}^{2} u(X),
\end{equation}
while it is independent from velocity perturbations, which therefore can be discarded.

This result allows a simplification in the choice of the physical situation: it is enough to discuss the case in which the condensate wavefunction has a constant phase, while its modulus slightly deviates from perfect homogeneity. It is convenient to introduce the parametrization:
\begin{equation}\label{eq:inhcondensate}
\psi = \left(\frac{\mu+\lambda}{\kappa}\right)^{1/2}(1+ u(\xx)),
\end{equation}
where $u(\xx)$ is a dimensionless function and it is assumed to be very small. In practice, we will assume that it is associated with a localized inhomogeneity of the condensate. At infinity we  ask that $u\rightarrow0$. Notice that the wavefunction \eqref{eq:inhcondensate} is real, due to the fact that we are really interested in number density fluctuations, and not on fluctuations in the velocity profile.
This simplification reduces the number of independent functions without making the system trivial, as it will be shown.
\subsection{The gravitational potential for the quasi-particles}\label{sec:gravpot}
Having discussed the setup in which we are working, we can reconsider the quasi-particle dynamics in this new system. It is not necessary to recover some sort of acoustic metric. While it is certainly interesting, given that the quasi-particle are massive and that we are interested in the nonrelativistic limit, the notion of acoustic metric is of little interest, in this particular case: it has been used just as a guide to isolate a candidate for the gravitational potential.

The first step, then, is to see if there is a term in the equation of motion for quasi-particles \eqref{eq:qpdynamics}, which can be identified as an external potential term. This will allow us to check the conjecture that it will be given by the number density perturbation $u(\xx)$, as well as the precise coefficient
relating it to the familiar Newtonian potential (having dimensions of the square of a velocity).
The next step will be to take this potential and plug it into the Bogoliubov--de Gennes equation \eqref{BdG}, which is describing its dynamics.

To identify the Newtonian potential, the diagonalization of the Hamiltonian in \eqref{eq:qpdynamics} for the field $\hchi$ must be done again, including now the fluctuations of the condensate wavefunction.
In this case, the diagonalization procedure is more involved: we have to deal with the non-commuting operators $\nabla^2$ and $u$.
We can not perform it in an exact way. However, we are interested in the Galilean regime for the quasi-particle spectrum, when $p_C\gg p$ {(cf section \ref{sec:regimes})}. It is then a reasonable approximation to neglect all the terms involving the commutators $[\hat{p}^2/2m,u(\xx)]$, which are largely suppressed (with respect to the other terms appearing in the equations) by the mass of the atoms and from the smallness of $u(\xx)$.

With these simplifying assumptions, the Hamiltonian for the quasi-particles in the non-relativistic limit is
\begin{equation}\label{eq:nonrelhquasip}
\hat{H}_{quasip.} \approx \mathcal{M} \csou^2- \frac{\hbar^2 \nabla^2}{2\mathcal{M}} + 2\frac{(\mu+\lambda)(\mu+4\lambda)}{\mathcal{M} \csou^2}u(\xx),
\end{equation}
where the mass of the quasi-particle $\mathcal{M}$ and for the speed of sound $c_{s}$ are given in \eqref{eq:soundspeedmass}. We first recognize the constant shift $\mathcal{M} \csou^2$ of the energy due to the rest mass in the Galilean regime. This term is not affecting the discussion in any way and can be subtracted without physical consequences. The  term proportional to $u(\xx)$ can be clearly interpreted as an external potential. If we want to  identify it with the gravitational potential $\phigrav$, we need to have
\begin{equation}\label{gravitational potential}
2\frac{(\mu+\lambda)(\mu+4\lambda)}{\mathcal{M} \csou^2}u(\xx)= \mathcal{M}\phigrav \Leftrightarrow \phigrav(\xx) = \frac{(\mu+4\lambda)(\mu+2\lambda)}{2\lambda m} u(\xx),
\end{equation}
where $\mathcal{M}$ is the mass of the quasi-particles. Note that this identification is  formal, and relies on the way in which the gravitational potential enters the Schroedinger equation for a non-relativistic quantum particle. We should always work with $u$: our definition of $\phigrav$ is dictated from the analogy we want to make with Newtonian gravity. For instance, we see that this definition becomes singular when we deal with massless quasi-particles, \ie when $\lambda\dr 0$. This must be expected: when $\lambda$ vanishes the quasi-particles become massless phonons, for which the coupling to a Newtonian gravitational potential cannot be defined in terms of their mass density.

\subsection{The modified Poisson equation}\label{sec:poisson}
Now that we have identified a candidate for the Newton potential $\phigrav$ from the quasi-particles dynamics, we need to check  that it satisfies some sort of Poisson equation. Since the gravitational potential is deduced from  $\psi$ -- as small deviations from perfect homogeneity (\cf \eqref{eq:inhcondensate}) -- the Poisson equation should be deduced from the BdG equation \eqref{BdG}. With the natural assumption that the potential  is reacting instantaneously to the change of distribution of matter, we can neglect the time derivative and \eqref{BdG} becomes
\begin{equation}\label{BdG1}
\left(\frac{\hbar^{2}}{2m} \nabla^{2} - 2 (\mu+\lambda) \right) u(\xx) = 2 \kappa \left( {\nnn}(\xx)+\frac{1}{2} {\mmm}(\xx) \right).
\end{equation}
We have seen in section \ref{sec:cond} that the terms ${\mmm}(\xx)$ and ${\nnn}(\xx)$ are functions of the atoms $\hat \chi$ outside the condensate  and therefore of the quasi-particle $\hat \phi$, through the Bogoliubov transformation \eqref{bogolobov transf}.  Therefore they can be interpreted as the source in the (modified) Poisson equation. We examined different types of source: either localized particles or plane-waves. We shall discuss here only the first kind of sources and forward the reader to \cite{gravito-BEC} for the second kind.

\subsubsection{Localized sources}\label{sec:loca-source}
The most natural source to consider for the Poisson equation  is  a single quasi-particle $\hat{\phi}$ at a given position $\xx_{0}$. However, point-like distributions give rise to divergencies. We consider therefore   a quasi-particle which is localized around the point $\xx_{0}$, with a non-zero spread to regularize these divergencies. We consider a quasi-particle in a state of the form:
\begin{equation}
\ket{\zeta_{\xx_0}} = \int \dv \xx \zeta_{\xx_0}(\xx) \hat{\phi}^{\dagger}(\xx)\ket{\Omega},\qquad \textrm{with} \qquad \int \dv\xx |\zeta_{\xx_0}(\xx)|^{2} = 1 \Leftrightarrow  \langle{\zeta_{\xx_0}|\zeta_{\xx_0}}\rangle=1.
\end{equation}
 $\zeta_{\xx_0}$ encodes the spreading of the particle around $\xx_{0}$ since
\begin{equation}
\bra{\zeta_{\xx_0}} \hat{\phi}^{\dagger}(\xx) \hphi(\xx) \ket{\zeta_{\xx_0}} = |\zeta_{\xx_0}(\xx)|^{2}.
\end{equation}
We can now determine the value for the  anomalous mass $\mmm$ and anomalous density $\nnn$ when the quasi-particle is in the state $\ket{\zeta_{\xx_0}}$.
An explicit calculation gives
\begin{eqnarray} \label{eq:locsource}
\nnn(\xx) = \left| \int \dv \zz f(\xx-\zz) \zeta_{\xx_0}(\zz) \right|^{2}  +\left| \int \dv \zz g(\xx-\zz) \zeta_{\xx_0}(\zz) \right|^{2} + \frac{1}{V} \sum_{\kkk} \beta^{2}(\kkk), \label{eq:anomalousdensity}\\
\mmm(\xx) = 2 \left(\int \dv \zz_{1} g(\xx-\zz_{1})\zeta^{*}_{\xx_0}(\zz_{1})\right) \left( \int \dv \zz_{1} f(\xx-\zz_{2})\zeta_{\xx_0}(\zz_{2})\right)
+ \frac{1}{V} \sum_{\kkk} \alpha(\kkk)\beta(\kkk)\label{eq:anomalousmass},
\end{eqnarray}
where we have introduced the functions $f$, $g$ depending on the Bogoliubov coefficients $\alpha$ and $\beta$
\begin{equation}
f(\xx) = \frac{1}{V} \sum_{\kkk} \alpha(\kkk)e^{i\kkk\cdot \xx}, \qquad g(\xx) = \frac{1}{V} \sum_{k} \beta(\kkk) e^{-i\kkk\cdot \xx}.
\end{equation}
The quantities $\nnn_{\Omega}$ and $\mmm_{\Omega} $  with
\begin{equation}
\nnn_{\Omega} = \frac{1}{V} \sum_{\kkk} \beta^{2}(\kkk), \qquad \mmm_{\Omega} = \frac{1}{V} \sum_{\kkk} \alpha(\kkk)\beta(\kkk),
\end{equation}
are vacuum contributions independent from the presence of actual quasi-particles. They are related to the inequivalence of the particle and quasi-particle vacua, and it can be easily seen that:
\begin{equation}
\nnn_{\Omega} = \bra{\Omega} \hchi^{\dagger}(\xx) \hchi(\xx) \ket{\Omega}, \qquad \mmm_{\Omega} = \bra{\Omega} \hchi(\xx) \hchi(\xx) \ket{\Omega}.
\end{equation}
The functions $f,g$ encode the fact that quasi-particles are collective degrees of freedom and therefore intrinsically some non-local objects. This non-locality is due to the Bogoliubov transformation \eqref{bogolobov transf}. Quasi-particles and atoms (\ie local particles) coincide only if we have $\alpha(\kkk)=1, \beta(\kkk)=0$, and therefore $f(\xx) = \delta^{3}(\xx)$,  while $ g(\xx)=0$. Since this is not the case, the anomalous mass and the anomalous density will show an intrinsic non-locality. The spreading characterized by $\ket{\zeta_{\xx_0}}$ encodes some extra non-local effect, introduced by hand for regularization purposes. Therefore this feature is not as fundamental as the non-locality introduced by the Bogoliubov transformation.

The  equation \eqref{BdG1} becomes then:
\begin{equation}\label{bdgbis}
\left(\frac{\hbar^{2}}{2m} \nabla^{2} - 2 (\mu+\lambda) \right) u(\xx) = 2 \kappa \left( \tilde{\nnn}(\xx)+\frac{1}{2} \tilde{\mmm}(\xx) \right) + 2 \kappa \left(
\nnn_{\Omega}+\frac{1}{2}{\mmm}_{\Omega} \right),
\end{equation}
 where we have  introduced the quantities
\begin{equation}
\tilde{\nnn}(\xx)=\nnn(\xx)-\nnn_{\Omega}, \qquad \tilde{\mmm}(\xx)=\mmm(\xx)-\mmm_{\Omega},
\end{equation}
which represent the contribution of actual quasi-particles to the anomalous density and anomalous mass, respectively.  By dimensional analysis, the terms $\nnn,\mmm$ have the dimensions of number densities. Since in Newtonian gravity the source for the gravitational field is a mass density, we introduce the mass density distribution:
\begin{equation}
\rho_{\rm matter} (\xx) = \mathcal{M}\left( \tilde{\nnn}(\xx)+\frac{1}{2} \tilde{\mmm}(\xx) \right).
\end{equation}
With this definition, we can rewrite \eqref{bdgbis} as an equation for the field $\phigrav$:
\begin{equation}\label{eq:modifiedpoisson}
\left( \nabla^{2} -\frac{1}{L^{2}} \right) \phigrav = 4 \pi G \rho_{\rm matter} + \Lambda,
\end{equation}
where we have defined
\begin{eqnarray}
&& G \equiv \frac{\kappa(\mu+4\lambda)(\mu+2\lambda)^{2}}{4 \pi \hbar^{2} m \lambda^{3/2} (\mu+\lambda)^{1/2}} ,\qquad
 \Lambda \equiv \frac{2 \kappa(\mu+4\lambda)(\mu+2\lambda)}{\hbar^{2} \lambda}(\nnn_\Omega +\demi \mmm_\Omega), \label{eq:lambda}\\
&& \label{eq:range}
L^{2} \equiv \frac{\hbar^{2}}{4m(\mu+\lambda)}.
\end{eqnarray}
This particular choice of notation is motivated by the comparison of \eqref{eq:modifiedpoisson} with the Newtonian limit of Einstein equations with a cosmological constant.
 %%%%%

For this reason, we can identify these three quantities as the analogous of the Newton constant, the analogous of the cosmological constant and a length scale which represents the range of the interaction, as we are going to discuss below.

To get a better grasp of the physics of the modified Poisson equation \eqref{eq:modifiedpoisson}, we can look at its solution for a given distribution of quasi-particle $\rho_{\rm matter}$.

As it is well known, a solution for the equation
\begin{equation}
\left( \nabla^{2} -\frac{1}{L^{2}} \right) \Phi(\xx) = 4 \pi G \mathcal{M} \delta^{3}(\xx-\zz),
\end{equation}
is given by the Yukawa potential
\begin{equation}
\Phi_{\rm Y} (\xx; \zz) = \frac{G \mathcal{M} e^{-|\xx-\zz|/L}}{|\xx-\zz|}.
\end{equation}
On the other hand, a solution for the equation
\begin{equation}
\left( \nabla^{2} -\frac{1}{L^{2}} \right) \Phi(\xx) = \Lambda,
\end{equation}
is just given by the constant solution
\begin{equation}
\Phi_{\Lambda} = - L^{2}\Lambda.
\end{equation}
Notice the peculiarity of this solution. It does not give rise to a gravitational acceleration since the gradient is trivially zero. Therefore the only effect of this term is to shift the overall density of the condensate.

The linearity of equation \eqref{eq:modifiedpoisson} allows us to use these results to write down a solution for a generic distribution of matter (\ie quasi-particles) as
\begin{equation}
\phigrav(\xx) = \int  \rho_{\rm matter}(\zz) \Phi_{\rm Y}(\xx;\zz) \dv \zz + \Phi_{\Lambda}.
\end{equation}
Solutions of \eqref{eq:modifiedpoisson} are therefore constructed from the Yukawa potential smeared out due to the non-locality of the quasi-particle (with an extra global shift due to the cosmological constant). The Yukawa potential is typically encoding some short range interaction, characterized by the scale $L$ which is  simply related to the healing length  \eqref{healing},
\be \label{eq:rangegrav} L^2=\frac{\xi^2}{2}.\ee
Although this a very short range for the gravitational interaction, this outcome should not come as a surprise.
In fact, the healing length (\cf \eqref{healing}) characterizes the typical length over which a condensate can adjust to density gradients.  Since density inhomogeneities encode the gravitational interaction, one should expect them to be damped over a distance of the order the healing length.

In the context of relativistic field theory, the short interaction scale for gravity would be translated in a  massive graviton, with mass given by $$M_{grav}^2=\frac{\hbar^2}{L^2\csou^2}=4\frac{\mu+\lambda}{\mu + 2 \lambda} m^{2}.$$
We can then compare the masses of the quasi-particles $\mathcal{M}$, graviton $M_{grav}$ and atoms $m$,
\begin{equation}
0\leq \mathcal{M} <m < \sqrt{2}m < M_{grav}\leq 2m,
\end{equation}
which shows the hierarchy of the energy scales present in this system. We notice that the graviton is then always more massive than the quasi-particles, and that this interaction is of very short range, since the $\xi$ is much shorter than the acoustic Compton length\footnote{We are using $c_{sound}$ instead of $c_{light}$ to define all these scales. We have to use the natural units for a hypothetical phononic observer.} of the quasi-particles. In particular, we cannot tune the parameters of the system in such a way to make $M_{grav}$ arbitrarily small, in order to be closer to reality.

\subsection{Lessons}
In an analogue gravity model based on a BEC system, the degrees of freedom are separated into  the atoms that condense and the ones which do not.  Quasi-particles are then collective degrees of freedom constructed from the non-condensed atoms. The dynamics of the quasi-particles can be described, in a given regime, in terms of the propagation of particles over an effective curved spacetime metric, which is a function of the density $n_c$ and the velocity profile $\vec v$ of the condensate. In this sense, it is natural to expect that gravitational degrees of freedom are encoded in the condensate. Dynamics of the latter is encoded in the BdG equation \eqref{BdG}, which is essentially Galilean. Hence, we can not expect to recover the Einstein equations in this context \cite{Barcelo:2001tb}. Nevertheless, one can still try to interpret  \eqref{BdG} as some sort of Poisson equation for some type of Newtonian gravity.

However, quasi-particles are massless in usual BEC systems and hence they cannot  be considered as sources for the gravitational field in the Poisson equation. We introduced therefore a new term $\hat H_\lambda$ in the dynamics of the BEC which softly breaks the $U(1)$ symmetry and consequently, as we showed in section \ref{sec:quasipart}, generates a mass gap for the quasi-particles. We showed explicitly that the presence of this small symmetry breaking term does not prevent a condensation from happening and still allows a mean field description (which is sufficiently accurate for our purposes). Then, following the usual general relativistic argument, we have argued, in section \eqref{sec:grav},  that the Newtonian potential $\Phi_N$ has to be  related to small inhomogeneities in the condensate density (while perturbations in the velocity profile do not contribute at first order as gravitational degrees of freedom). This conjecture, based on the analysis of a standard BEC system, was then confirmed by a specific analysis of the modified BEC dynamics for an almost homogenous condensate.

The end point of this investigation can be then summarized in the following two equations
\begin{eqnarray}
&& {\vec{F}} = {\mathcal{M}} \vec{a} = - {\mathcal{M}} \vec{\nabla} \phigrav,\\
&& \left(\nabla^2-\frac{1}{L^2} \right) \phigrav = 4 \pi G_N \rho + \Lambda,\label{poisson}
\end{eqnarray}
where ${\mathcal{M}}$ is the mass of the quasi-particle acquired via the soft $U(1)$ symmetry breaking induced by \eqref{sy-br}, $L$ is proportional to the healing length,  $\Lambda$ plays the role of the cosmological constant and $G_N$ is an effective coupling constant that depends on the condensate microphysics and the form of the matter source.

For what regards the latter we have considered two cases: a localized quasi-particle state and a set of plane waves. In the first case the analogue Newton constant is indeed momentum and position independent and the solution of the modified Poisson equation  \eqref{poisson},  has the form of a smeared Yukawa potential. The smearing is due to the fact that quasi-particles are intrinsically non-local objects, being collective degrees of freedom. When considering plane-waves as sources, we have instead that, due to the momentum dependence of the Bogoliubov transformation, $G_N$ is running with the momentum and the solution for the gravitational potential is a constant (albeit a different one for different momenta). One should however be careful: while it is common in quantum field theory (QFT) to encounter the notion of running coupling constants, the origin of the running here is rather peculiar. Indeed, in QFT the running is due to quantum corrections to the tree level/classical action, here the running is due to the inequivalence between the ground state of the Fock spaces of atoms and quasi-particles. Paraphrasing what has been done in the context of emergent geometry, where the notion of ``rainbow geometry" has been introduced, we could speak about ``rainbow dynamics''.

We have also obtained naturally a cosmological constant in the model: vacuum gravitates, even though in a very peculiar way. It  is induced by the terms $\bra{\Omega}\hchi^{\dagger}\hchi\ket{\Omega},\bra{\Omega}\hchi\hchi\ket{\Omega}$, where $\Omega$ is the state with no quasi-particles. It is entirely due to the (unavoidable) inequivalence between the quasi-particle vacuum and the particle vacuum and cannot be put to zero just tuning the parameters. It represents an interesting alternative to known mechanisms to generate a cosmological constant (see also \cite{L-Vol} for similar ideas about the nature of the vacuum energy in condensed matter systems).

Let us compare this point with the standard cosmological constant problem. If one thinks to the cosmological constant is generated by the zero point energy associated to each mode, in the case of BECs a naive expectation would be that the cosmological constant term would be set at the characteristic UV scale represented by the inverse of the healing length. This would lead to a very large vacuum energy. However, we have seen that the cosmological constant, in this specific case, is linked to the so called depletion factor (\ie the ratio between the non-condensed fraction and the condensed one), and hence naturally suppressed by a factor $1/N$ in the expansion in the inverse of the number of atoms. It is the condensation mechanism itself guaranteeing a naturally small cosmological constant. Of course, the worse is the condensation, the larger is the number of atoms out of the condensate and hence the bigger will be the cosmological constant. This fact has a rather peculiar interpretation.

To have an emergent Lorentzian structure, it is essential that there is a mean field which represents the emergent spacetime. This was encoded in the splitting of the field operators $\hpsi \approx \psi + \hchi$. However, for a mean field approach to make sense, fluctuations around the mean field should be relatively small:
\begin{equation}
\frac{\langle \hchi^{2} \rangle}{ |\psi|^{2}} \ll 1.
\end{equation}
Therefore, the smallness of the analogue of the cosmological constant term in the BEC is deeply intertwined with the very definition of mean field, \ie how good is the picture of quasi-particles moving in a classical Lorentzian spacetime. Of course this poses the question of whether this mechanism do apply also for spacetime in which we actually live.

In conclusion, BEC as an analogue model for gravity presents many differences with a realistic gravity theory as we expected. We do not get general relativity in a condensate. However

\begin{itemize}
\item there is an emergent Lorentzian metric describing the propagation of the phonons;
\item there is a mismatch between microlocality and macrolocality due to the nonlocal nature of the phonons;
\item there is an emergent Newtonian gravitational theory, which is very short range;
\item in this theory vacuum gravitates;
\item the source term for the gravitational field inherits the nonlocality properties of the phonons;
\item the cosmological constant is naturally small provided that the depletion factor is small.
\end{itemize}

Despite the limited applicability of the results, the suggestions for realistic theories of gravity, in particular for quantum gravity and the role of locality, and, perhaps more interestingly, for the cosmological constant problem are definitely worth of further investigation.

\section{Emergent gravity: the role of symmetries}

The discussion of the BEC model has shown that by limiting the analysis to condensed matter systems there are rather strong constraints on the kind of gravitational models which is reasonable to simulate. For example, in the above investigation we had a single scalar field: it would be interesting is to see what happens if several different species are present. In that case, besides the issue of having a short range rather than a long range interaction, also the coupling to the gravitational field must be carefully discussed. Indeed, in order to have some sort of equivalence principle, all the fields must be coupled to the gravitational field in the same way.
The natural setup to discuss these issues is the 2-BEC model \cite{Liberati:2005id,Weinfurtner:2006wt}: in fact in this case one could treat a multi-particle system whose richness could allow a closer mimicking of Newtonian gravity with a long range potential.
However, the fact that emergent gravity has to be Newtonian in a BEC-based analogue model seems to be unavoidable since the gravitational potential depends on the condensate, which is typically described by non-relativistic equations. A possible way to avoid this issue is either to consider relativistic BEC \cite{Rel-BEC,Rel-BEC2} (however in this case we would still expect to get only some type of scalar gravity), or to change completely paradigm and identify gravity not as the condensate but  as linked, together with matter, to the perturbations around the condensate. We will consider later this second point of view in a different model.

Furthermore, there is another important issue that needs attention. In our treatment we neglected the quantum potential, i.e. we have deliberately worked in the hydrodynamic limit of the theory, carefully avoiding the issue of the breakdown of acoustic Lorentz invariance in the system at suitably high energies of the quasi-particles. Presumably the breakdown of this emergent spacetime symmetry, namely local Lorentz invariance, will be linked also to some relevant regime change in the gravitational dynamics (which is anyway affected by the presence of a Lorentz symmetry breaking scale, the healing length, which ends up setting the graviton mass scale). Should we take Lorentz symmetry breaking as a crucial ingredient of the emerging gravity paradigm or as an accident of the condensed matter analogue models? In the first case, how the breakdown of such spacetime symmetry affects the symmetries of the gravitational dynamics and in particular diffeomorphism invariance? Furthermore, does this imply that an emergent gravity scenario should give up the relativity principle and bring us back to Newton's absolute space and time?

In order to explore these issues we can start investigating the role of Lorentz invariance in emergent gravity scenarios by considering the most well known  ``no-go theorem" against them, i.e. the so called Weinberg--Witten theorem  \cite{WW}.

\subsection{LIV and emergent gravity: the Weinberg--Witten theorem}\label{sec:ww}

The idea of having the graviton as a composite particle/emergent field is certainly a fascinating idea. However, there are limitations to what it is possible to do. In particular, there is a theorem, due to Weinberg and Witten \cite{WW}, which is often presented as a crucial (fatal, in fact) obstruction for a successful emergent gravity program.

The theorem states precise limits for the existence of consistent theories with massless particles. It has two parts, and it says that (quoting from \cite{WW}):
\begin{quote}\begin{enumerate}
\item A theory that allows the construction of a Lorentz-covariantly conserved four-vector current $J^{\mu}$ cannot contain massless particles of spin $j>1/2$ with nonvanishing values of the conserved charge $\int J^{0}d^{3}x$.
\item A theory that allows for the construction of a conserved Lorentz covariant energy-momentum tensor $\theta^{\mu\nu}$ for which
$\int \theta^{0\nu}d^{3}x$ is the energy-momentum four-vector cannot contain massless particles of spin $j>1$.
\end{enumerate}
\end{quote}

For a careful discussion of the proof of the theorem, and for references, see \cite{Loebbert}. For additional comments, see \cite{Jenkins1,Jenkins2}.

Crucial ingredients for the proof of this theorem are Lorentz invariance and the nonvanishing of the charges obtained from Lorentz covariant vectors and tensors. Interestingly, the gauge bosons like the gluons and the graviton are not forbidden since the current for the gluons is not Lorentz-covariant conserved, and the graviton does not possess a covariant stress-energy tensor (but rather a pseudo-tensor).

This theorem, then, poses rather strong constraints on the possible theories that can be built in Minkowski spacetime. Of course, gravity is not just the theory of a spin-2 particle in Minkowski spacetime. Nevertheless, it surely makes sense to consider the linearized theory in sufficiently small neighborhoods. In this limit, then, the theorem does apply.

With this caveat in mind, we can say that in an emergent gravity program this theorem must be taken appropriately into account and appropriately evaded.
There are (at least) two ``obvious'' way out:
\begin{itemize}
\item allow for Lorentz symmetry breaking, or
\item make the spacetime manifold to emerge as well.
\end{itemize}

The first option is rather straightforward, and it is essentially what could be pursued within scenarios like the one considered in analogue models, in which a preferred time function is specified. However, there is apparently also a (conceptually high) price to pay: a step back from Minkowski spacetime to the notions of absolute space and time. Moreover, and most importantly, there is the issue of recovering a low energy approximate Lorentz invariance. We shall come back later on these issues.

The second option is probably the most viable, conceptually appealing, but most demanding in terms of new concepts to be introduced. If no reference is made to a background Minkowski spacetime, but rather the graviton emerges in the same limit in which the manifold emerges, then there is no obvious conflict with the Weinberg-Witten theorem. Simply, what is called the gauge symmetry in terms of fields living of spacetime is the manifestation of an underlying symmetry acting on the fundamental degrees of freedom in the limit when they are reorganized in terms of a spacetime manifold and fields (gauge fields and gravitons in particular).
There are already two examples of this possibility, namely matrix models and quantum graphity models. In both cases, the very notion of spacetime manifold is immaterial for the foundations of the theory. The manifold and the metric are derived concepts, obtained in precise dynamical regimes of the theory. The interested reader can find additional comments and references in \cite{Konopka:2008hp,MatrixModels}.

The bottom line of this very concise overview of the WW theorem is clear: to obtain a realistic model of emergent gravity one must ask for very special mechanisms to be at work in the model. Without these, the theory would not be able to give a meaningful limit.

\subsection{Why breaking Lorentz symmetry might be good}

Specific hints of LV arose from various approaches to Quantum Gravity.
Examples include string theory tensor VEVs~\cite{KS89}, spacetime foam~\cite{GAC-Nat}, semiclassical spin-network calculations in Loop QG~\cite{LoopQG}, non-commutative geometry~\cite{Carroll:2001ws, Lukierski:1993wx, AmelinoCamelia:1999pm}, some brane-world backgrounds~\cite{Burgess:2002tb} and condensed matter analogues of ``emergent gravity''~\cite{LRR,Analogues}. Although none of these calculations proves  that Lorentz symmetry breaking is a necessary feature of Planck scale physics, they did stimulate research aimed at understanding the possible measurable consequences of LV \cite{David,GAC-Rev,LibMac}. Furthermore, recent investigations strongly suggested that an high energy breakdown of Lorentz invariance might strongly improve the renormalizability of field theories~\cite{anselmi,Visser:2009fg} including gravitation~\cite{Horava:2009uw}.

This seems in close analogy with what we naively expect from analogue models scenarios like the BEC one. There, in fact, the renormalizability of the fundamental field theory, a non-relativistic $\lambda^4$ action, is preserved by the Bogoliubov transformation leading to the quantum field theory of the quasi-particles. Should we take these results as a strong hint that Lorentz symmetry breaking should be a part of any working emergent gravity scenario? It is at this stage unclear if we can be that bold. Surely one open issue is  the naturalness of theories endowed with Lorentz symmetry breaking. In general, radiative correction, or alternatively renormalization group running, lead to a dangerous ``percolation" in the infrared regimes of the Lorentz breaking \cite{Collins:2004bp,JLMAnnals,Iengo:2009ix}, something strongly constrained by current observations \cite{David,GAC-Rev,LibMac}. It seems that in order to solve this issue some sort of custodial symmetry would be needed (see e.g. the related discussion in \cite{JLMAnnals}) but no satisfactory solution up to date has been found.

In addition to the above, more technical, issue it is also clear that a more philosophical one is also present. As we said, many find quite unappealing the idea to give up the relativity principle and go back to a preferred system of reference. In this sense, however, some comments are in order on the relation between Lorentz invariance and relativity.

\subsection{Relativity beyond Lorentz?}

Lorentz invariance of physical laws relies on only few assumptions: the principle of relativity, stating the equivalence of physical laws for non-accelerated observers, isotropy (no preferred direction) and homogeneity (no preferred location) of space-time, and a notion of precausality, requiring that the time ordering of co-local events in one reference frame be preserved \cite{ignatowsky,Liberati:2001sd,Sonego:2008iu}.
In this sense a breakdown of Lorentz invariance does not necessarily imply a breakdown of the relativity principle. For this reason, it is worth exploring an alternative possibility that keeps the relativity principle but that relaxes one or more of the above postulates. Such a possibility can lead to the so-called very special relativity framework \cite{Cohen:2006ky}, which was discovered to correspond to the break down of isotropy and to be described by a Finslerian-type geometry~\cite{Bogoslovsky:2005cs,Bogoslovsky:2005gs,Gibbons:2007iu}. In this example, however, the generators of the new relativity group number fewer than the usual ten associated with Poincar\'e invariance. Specifically, there is an explicit breaking of the $O(3)$ group associated with rotational invariance.

One may wonder whether there exist alternative relativity groups with the same number of generators as special relativity. Currently, we know of no
such generalization in {(commutative)} coordinate space. However, it has been suggested that, {in non-commutative spacetime}, such a generalization is possible, and
it was termed  ``doubly" or ``deformed" (to stress the fact that it still has 10 generators) special relativity, DSR~\cite{ACDSR}. Unfortunately, the various DSR candidates  face in general major   problems regarding  their physical interpretation (e.g.~the so called ``soccer ball'' problem~\cite{ACDSR})}.

Finally, it is a {logical, and rather simple, possibility that  a Lorentz symmetry breakdown} could be signaling an interpolation from a relativity group to another one, for example two special relativity groups characterized by different limit speeds or between a Lorentzian and an Euclidean Poincar\'e group. This second possibility is quite appealing because it would allow to consistently introduce a minimum length without reducing the generators of the Poincar\'e group as well as it would give a natural meaning to the breakdown of Lorentz invariance as we know it by associating it with the emergence of time.

In what follows we shall pursue this route and try to built up a model of emergent gravitational dynamics where time and gravity will emerge from a Euclidean manifold endowed with a set of fields.

\section{Emerging time and scalar gravity}

The discussion of the BEC model has shown that by limiting the analysis to condensed matter systems there are rather strong constraints on the kind of gravitational models is reasonable to simulate. Therefore, we will leave the arena of analogue models and we will present a toy model in which a number of different issues can be addressed. In particular, we will focus on two of them.

First of all, given that analogue models are condensed matter systems, the notion of time is inherited from the time of the laboratory. This structure percolates from the fundamental level of Galilean spacetime where the atomic dynamics takes place onto the effective dynamics of the propagating degrees of freedom, \eg phonons.

There is an exception to this: by a careful tuning of the coupling constants, it is possible to make the scattering length of a BEC a negative quantity, and hence produce an effective dynamics for phonons which is Euclidean and not Lorentzian. Of course, this is a very interesting phenomenon which gives the possibility of studying a rather exotic class of phenomena related to signature change events~\cite{Weinfurtner:2007dq,White:2008xr}. Nonetheless, this phenomenon is rather simple, from a conceptual point of view: while time is present at the fundamental level as a definite structure (\ie a preferred class of foliation of the spacetime manifold), this latter is hidden for the emergent system.

It would be much more interesting to understand how it is possible to do the opposite, \ie whether it is possible to emerge time out of a timeless system. We will show a simple toy model in which this happens \cite{timenordstrom}.

The second point concerns the issue of diffeomorphism invariance. Diffeomorphism invariance is one of the distinctive features of General Relativity. In an emergent gravity program, it is crucial to understand how it will be possible to emerge it as well. By elaborating further on the toy model used to emerge time, we will show how a diffeo-invariant theory, namely Nordstr\"om theory for scalar gravity, can be extracted from this particular example.

\subsection{Emergence of time}

Let us assume that a fundamental unknown theory gives rise in some large number limit to  simple structures such as $\R^4$ equipped with the Euclidean metric $\delta^\mn$, and a set of scalar fields $\Psi_i(x_\mu)$, $i=1,...,N$ ($x_\mu\in\R^4$) with their Euclidean Lagrangian $\lll$. Since we do not know this fundamental theory, we choose such Lagrangian to be of the simple shape\footnote{We could also consider  a dependence on crossed terms of the kind $h^{\mn}\partial_\mu\Psi_i\partial_\nu\Psi_j$, however this is not changing the final result.}
\begin{equation}\label{fundamental Lagrangian}
\lll=F(X_1,..., X_N).
\end{equation}
with $X_i=\delta^{\mn}\partial_\mu\Psi_i\partial_\nu\Psi_i$.
It is easy to see that this Lagrangian is invariant under the Euclidean group $ISO(4)$.
The equations of motion are then simply for a given field $\Psi_i$
\begin{equation}\label{fundamental EOM}
 \partial_\mu \left(\frac{\partial F}{\partial X_i}\partial^\mu\Psi_i\right)=0=\Sigma_j{\left(\frac{\partial^2F}{\partial X_i\partial X_j}\partial_\mu X_j\right)} \partial^\mu\Psi_i + \frac{\partial F}{\partial X_i}\partial_\mu\partial^\mu\Psi_i.
\end{equation}

Let us now consider a specific solution of the above equations of motion, $\psi_i$ and perturbations $\varphi_i$ around it. For $\Psi_i=\psi_i+\varphi_i$, the kinetic term $X_i$ becomes then
\begin{equation*}
X_i\dr \ov{X}_i+\delta X_i, \quad \textrm{ with } \quad \ov{X}_i= \delta^{\mn}\partial_\mu\psi_i\partial_\nu\psi_i \quad \textrm{ and }
\end{equation*}
\begin{equation}  \delta X_i= 2\partial_\mu\psi_i\partial^\mu\varphi_i +\partial_\mu\varphi_i\partial^\mu\varphi_i.
\label{perturbations}
\end{equation}
We intend now to identify some specific $F$ such that the Lagrangian for the  perturbations $\varphi_i$ is invariant under the Poincar\'e group $ISO(3,1)$. To determine the Lagrangian for the perturbations $\varphi_i$, we expand \eqref{fundamental Lagrangian} using \eqref{perturbations}.
\begin{eqnarray}\label{expansion}
F(X_1, .., X_N)\dr && F(\ov{X}_1, ..,\ov{X}_N)+
\sum_j\left.\frac{\partial F}{\partial X_j }\right|_{\ov{X}}\delta X_j \nn \\
&&+\demi \sum_{jk}\left. \frac{\partial^2 F}{\partial X_j\partial X_k }\right|_{\ov{X}} \delta X_j\delta X_k +
 \frac{1}{6}\sum_{jkl}\left.\frac{\partial^3 F}{\partial X_j\partial X_k \partial X_l}\right|_{\ov{X}}\delta X_j\delta X_k\delta X_l+ ...
\end{eqnarray}
The first term $F(\ov{X}_1, .., \ov{X}_N)$ is the Lagrangian for the classical solution $\psi_i$. The second term, the one linear in $\delta X_j$, contains a term linear in $\partial_\mu\varphi_i$, which is zero on shell. We can also identify the quadratic  contribution for $\partial_\mu \varphi_k\partial_\nu \varphi_k$:
\begin{eqnarray}
&& \textrm{for } k\neq l, \quad \partial_\mu \varphi_k \partial_\nu \varphi_l \left(  2 \left. \frac{\partial^2 F}{\partial X_k\partial X_l }\right|_{\ov{X}}\partial^\mu \psi_k\partial^\nu \psi_l\right),\label{crossed terms}\\
&& \textrm{for } k= l, \quad \partial_\mu \varphi_k \partial_\nu \varphi_k \left( \left. \frac{\partial F}{\partial X_k }\right|_{\ov{X}}\delta^{\mn} + \demi \left. \frac{\partial^2 F}{(\partial X_k)^2 }\right|_{\ov{X}}\partial^\mu \psi_k\partial^\nu \psi_k\right) \label{diagonal terms}.
\end{eqnarray}
The contribution \eqref{crossed terms} introduces some mixing between fields in the kinetic term. To simplify the analysis, we demand  that they cancel, which puts a constraint on the choice of $F$, \ie $ \left. \frac{\partial^2 F}{\partial X_k\partial X_l }\right|_{\ov{X}}=0$, if $k\neq l$. A specific solution is then
\begin{equation}\label{first cond for F}
F(X_1, .., X_N)= f_1(X_1)+...+f_N(X_N).
\end{equation}
We can identify in \eqref{diagonal terms} the effective or emergent metrics\footnote{Actually, we show here the inverse metrics from which the actual metrics can be derived once invertibility conditions are imposed. In our case of interest, this will always be true.} for each field $\varphi_k$, (taking into account \eqref{first cond for F})
\begin{equation}
g^\mn _k\equiv \left. \frac{d f_k}{d X_k }\right|_{\ov{X}_k}\delta^{\mn} + \demi \left. \frac{d^2 f_k}{(d X_k)^2 }\right|_{\ov{X}_k}\partial^\mu \psi_k\partial^\nu \psi_k.
\label{eff-metr}
\end{equation}
Since a priori $f_i\neq f_j$ and $\psi_i\neq \psi_j$ if $i\neq j$, we are dealing with a multi-metric structure: each field sees its own metric. However, we can enforce a mono-metric structure
by constraining the solution $\psi_k$ and the derivatives of $f_k$ at $\ov{X}_k$ to be independent of $k$
\begin{equation}\label{second con for F}
f_k=f, \quad \psi_k=\psi, \quad \forall k.
\end{equation}

So far we have just shown that the perturbations around a solution of the field equations on a Riemannian manifold can propagate, for suitably chosen Lagrangians, on an effective geometry which is not the fundamental one, $\delta_{\mu\nu}$, but rather a rank 2 tensor constructed from it and partial derivatives of the chosen background solution.  Note that, in order for this to be possible, it was crucial to have a starting Lagrangian with non-canonical kinetic terms as it can be clearly evinced by the second contribution to the metrics in equation \eqref{eff-metr}.  As a next step, we show now how for some solutions of the equations of motion, such effective metric can be of pseudo-Riemannian form. In fact, we can even ask that the metric (\ref{eff-metr}) is the Minkowski metric $\eta_\mn$. This will put some constraints on the derivative of $f$, evaluated at $\ov{X}=\partial^\mu \psi\partial_\mu \psi$.

In order to do so, we shall need to specify a particular solution, $\bar{\psi}$, of the equations of motion. Let us take it to be an affine function of the coordinates, $\bar{\psi}= \alpha^\mu x_\mu + \beta$. It is easy to check that this is indeed a solution of our field equations \eqref{fundamental EOM}. Moreover, thanks to the $SO(4)$ symmetry, we can always make a rotation such that
\be\label{choice of psi} \bar{\psi}= \alpha x_0 + \beta.\ee
The choice of the coordinate $x_0$ is completely arbitrary, what only matters is that there is one coordinate which is pinpointed.  Finally, we ask for the metric to have the signature $(-,+,+,+)$.  This puts some constraint on the value of  the derivatives of $f$
\begin{eqnarray}
&&\left. \frac{df}{d X }\right|_{\ov{X}} + \demi \left. \frac{d^2 f}{(d X)^2 }\right|_{\ov{X}}\partial^0 \bar{\psi}\partial^0 \bar{\psi}<0,\nn\\
&& \left. \frac{d f}{d X }\right|_{\ov{X}} + \demi \left. \frac{d^2 f}{(d X)^2 }\right|_{\ov{X}}\partial^a \bar{\psi}\partial^a \bar{\psi}>0, \quad a=1,2,3\label{pre-Lor-cond}
\end{eqnarray}
which using \eqref{choice of psi} imply
\begin{equation}
\left. \frac{df}{d X }\right|_{\ov{X}}+\frac{\alpha^2}{2}\left. \frac{d^2 f}{(d X)^2 }\right|_{\ov{X}}<0, \qquad  \left. \frac{df}{d X }\right|_{\ov{X}}>0.
\label{Lor-cond}
\end{equation}
Note that, due to the choice of a solution of the form \eqref{choice of psi}, the conditions \eqref{pre-Lor-cond} are not only implying a pseudo-Riemannian signature but also the constancy of the metric components, which hence can be easily rescaled so to take the familiar Minkowskian form ${\rm diag}(-1,+1,+1,+1)$.

Of course, there are many possible choices of $f(X)$ and $\alpha$ which can fulfill the above requirements.
For example, we can pick up the specific combination
\begin{equation}\label{solution for f}
f(X)= -X^2+ X,  \qquad \frac{1}{3}<\alpha^2<\frac{1}{2}.
\end{equation}
However, in what follows we should not make use of any particular form of $f(X)$ and $\alpha$ and simply assume that they are such that \eqref{Lor-cond} are satisfied.

To summarize, since $g^\mn _k\equiv \eta^\mn$, $\forall k$, the (free) perturbations $\varphi_i$ are propagating on a Minkowski space, even though the fundamental theory  is Euclidean (\cf \eqref{fundamental Lagrangian}).
At this point few remarks are in order.

So far, our theory does not posses any fundamental speed scale. This is natural since the fundamental theory is Euclidean. At this level, there is no coordinate with time dimension and therefore one cannot define a constant with speed dimension. The invariant speed $c$, which will relate the length $x_0$ to an actual time parameter $t$, could be determined experimentally by first introducing a coordinate with time dimension (as it would be natural to do given the hyperbolic form of the equations of motion for the perturbations)  and then by defining $c$ as the signal speed associated to light cones in the effective spacetime\footnote{
Noticeably, a similar situation is encountered in the von Ignatowsky derivation of Special Relativity \cite{ignatowsky} where, given a list of simple axioms, one derives the existence of a universal speed, observer independent, which is not fixed a priori to be the speed of light but has to be identified via actual experiments.}.

Second, a comment is due about our choice of the background solution around which we have considered the dynamics for perturbations. It is obvious that within our model this choice is arbitrary. It simply shows that there are some background solutions $\bar{\psi}$ for which a pseudo-Riemannian metric can emerge. Obviously, different background solutions could lead to alternative metrics, \eg one could also obtain the Euclidean metric $\delta_\mn$ (for example if $\psi$ is constant), a degenerate metric or more complicated structures according to the possible solutions $\psi$. While it is conceivable that in a more complicate model we could have some mechanism for selecting the specific background solution that leads to an emergent Lorentzian signature, it is not obvious at all that such a feature should be built in the emergent theory. In fact, one generally minimizes an energy functional to select the ground state of the theory. However, when looking at Lorentzian signature emergence starting from an Euclidean set up as in our model, there is no initial notion of time and hence no energy functional to minimize. It is therefore unclear how a ground state could be selected from within the emergent system.

 On the other hand, it is also conceivable that the actual background solution in which the initial system of fields (\ref{fundamental Lagrangian}) emerges from the fundamental (pre-manifold) theory, can be
 depending on
the conditions for which the ``condensation" of the fundamental objects takes place. In this sense,  the right ground state or background solution would be selected from minimizing some functional defined  at the level of the atoms of space-time. To use an analogy, the same fundamental constituents, \eg carbon atoms, can form very different materials, diamond or graphite, depending on the external conditions during the process of formation. Similarly, in a Bose--Einstein condensation the characteristics of the background solution (the classical wave function of the condensate), such as density and phase, are determined by physical elements (like the shape of the EM trap or the number and kind of atoms involved) which pre-exist the formation of the condensate.

In conclusion, we have identified the fundamental Lagrangian so that the  perturbations $\varphi_i$ have a kinetic term determined by the Minkowski metric.
\begin{equation}\label{Lagrangian for perturbations}
\lll_{\rm eff}(\varphi_1,...\varphi_N )= \sum_i  \eta^{\mn} \partial_\mu\varphi_i\partial_\nu\varphi_i.
\end{equation}
In this sense, we have a toy-model for the emergence of the Poincar\'e symmetries.
This construction can be seen as a generalization of the typical situation in analogue models of gravity \cite{LRR} where one has Poincar\'e symmetries emerging from fundamental Galilean symmetries \cite{LRR}. However, let us stress that in our case no preferred system of reference is present in the underling field theory given that the fundamental Lagrangian is endowed with a full Euclidean group $ISO(4)$.
Moreover, the emergence of a pseudo-Riemannian metric is in our model free of the usual problems encountered in the context of continuous signature change (\eg degenerate metrics) given that the former arises as a feature of the dynamics of perturbations around some solution of the equations of motion. Similarly one can see that  the invariance under Lorentz transformations is only an approximate property of the field equations (as usual for emergent systems), valid up to some order in perturbation theory.  In particular, if we analyze the third order contribution in \eqref{expansion} we get\footnote{We are in the mono-metric case, so that $F(X_1,..., X_N)=f(X_1)+...+f(X_N)$, and  $\psi_k=\psi$, $\forall k$.}
 \begin{equation}\label{cubic contribution}
\partial_\alpha \varphi_k \partial_\beta \varphi_k\partial_\gamma \varphi_k \left(\left. \frac{d^2 f}{(d X_k)^2}\right|_{\ov{X}}\partial^\alpha\psi \delta^{\beta\gamma}+ \frac{1}{6}\left. \frac{d^3 f}{(d X_k)^3}\right|_{\ov{X}}(\partial^\alpha \psi \partial^\beta \psi \partial^\gamma \psi ) \right).
\end{equation}
This contribution is clearly not Lorentz invariant if the solution $\psi$ pinpoints a specific direction,  as for example when the Minkowski metric is emergent. As a matter of fact our theory will show {\ae}ther like effects beyond second order.

So far, we have hence generalized and extended results familiar to the analogue gravity community. However, as said, a typical drawback of analogue gravity models is related to the fact that they show only the emergence of a background Lorentzian geometry while they are unable to reproduce a geometrodynamics of any sort. In what follows, we shall show that our model overcomes this drawback and indeed is able to describe the emergence of a  theory for scalar gravity. This theory will come out to be the only known other theory of gravitation, apart from General Relativity, which satisfy the strong equivalence principle~\cite{Will}, \ie Nordstr\"om gravity.

\subsection{Emergence of Nordstr\"om gravity}

In this section, we describe how we can recover a relativistic scalar gravity theory from a Lagrangian of the type \eqref{fundamental Lagrangian}, when ground state is such that  the perturbations are living (at the lowest order in perturbation theory) in a Minkowski spacetime.  So, let us start from the truncated Lagrangian for the perturbations \eqref{Lagrangian for perturbations} that we obtained in the previous section. This Lagrangian can simply be  rewritten in terms of the (real) multiplet $\varphi= (\varphi_1,...,\varphi_N)$ as
\be \label{Lagrangian for perturbations 1}
\lll_{\rm eff}(\varphi)= \eta^\mn (\partial_\mu  \varphi)^T (\partial_\nu  \varphi).
\ee
This system has a global $O(N)$ symmetry which has emerged as well from the initial Lagrangian \eqref{fundamental Lagrangian}. It is hence quite natural to rewrite  the multiplet $\varphi$  by
introducing an amplitude characterized by a scalar field $\Phi(x)$  and a multiplet $\phi(x)$ with $N$ components such that\footnote{Our field redefinition is the generalization of the so-called Madelung representation \cite{LRR}.}
\begin{equation} \label{field redefinition}
\left(\begin{array}{c}\varphi_1\\\vdots
\\\varphi_N\end{array}\right)= \Phi\left(\begin{array}{c}\phi_1\\\vdots
\\\phi_N\end{array}\right),  \qquad \textrm{with } |\phi|^2\equiv\sum_i \phi_i^2=\ell^2.
\end{equation}
$\ell$ is an arbitrary length parameter to keep the dimension right. In particular, $\Phi$ is dimensionless and $\phi$ has the dimension of a length. $\Phi$ is the field invariant under $O(N)$ transformations, whereas  $\phi$ does transform under $O(N)$. As we shall see, this  field redefinition will provide us the means to identify gravity and matter degrees of freedom. The Lagrangian for the perturbations \eqref{Lagrangian for perturbations 1} reads now as\footnote{We use the normalization condition $|\phi|^2=\ell^2$, which implies in particular $\sum_i\phi_i \partial_\mu \phi_i=0$.}
\begin{equation*}
\lll_{\rm eff}(\varphi_1,...\varphi_N )\dr \lll_{\rm eff} (\Phi,\phi_1,...\phi_N)=
\end{equation*}
\begin{equation}
=\ell^2 \eta^{\mn} \partial_\mu\Phi\partial_\nu\Phi + \sum_i\Phi^2 \eta^{\mn}\partial_\mu\phi_i\partial_\nu\phi_i + \lambda(|\phi|^2-\ell^2),
\label{new Lagrangian}
\end{equation}
where $\lambda$ is a Lagrange multiplier. We recognize  in particular  the action for a non-linear sigma model given in terms of the fields $\phi_i$. The associated equations of motion are
\begin{eqnarray}
&& \eta^{\mn} (\ell^{2}\partial_\mu\partial_\nu\Phi - \Phi \sum_i\partial_\mu\phi_i\partial_\nu\phi_i) =0, \label{gravity EOM}\\
&& \eta^{\mn} (2\partial_\mu\Phi\partial_\nu\phi_i + \Phi^2 \partial_\mu\partial_\nu\phi_i+ \frac{1}{\ell^{2}} \partial_{\mu}\phi_{j}\partial_{\nu}\phi_{k}\delta^{jk} \phi_{i}) =0, \label{matter EOM}\\
&& |\phi|^2-\ell^2=0. \label{constraint}
\end{eqnarray}
If we introduce the (conformally flat) metric
\be\label{conformal-metric 1}
g_{\mn}(x)= \Phi^2(x)\eta_{\mn},
\ee
the equations of motion \eqref{matter EOM} can be simply rewritten as
\begin{equation*}
(\sqrt{-g})\mone\partial_\mu(\sqrt{-g} g^{\mn} \partial_\nu \phi_i)+ \frac{1}{\ell^{2}} g^{\mu\nu} \partial_{\mu}\phi_{j}\partial_{\nu}\phi_{k}\delta^{jk} \phi_{i}=
\end{equation*}
\begin{equation}\Box_g \phi_i +\frac{1}{\ell^{2}} g^{\mu\nu} \partial_{\mu}\phi_{j}\partial_{\nu}\phi_{k}\delta^{jk} \phi_{i}=0,
\label{matter}
\end{equation}
where we have introduced the d'Alembertian $\Box_g$ for the metric $g$ and used that $ \sqrt{-g}=\Phi^4$ and $g^{\mn}= \Phi^{-2}\eta^{\mn}$. Notice that equation \eqref{matter EOM}  can be rewritten in the form \eqref{matter} using the metric redefinition \eqref{conformal-metric 1} only in four dimensions.
To be consistent, the change of variable  $\Phi\dr g_\mn$ should be completed with the constraint that $g_\mn$ is conformally flat, that is
\begin{equation}\label{weyl tensor}
C_{\alpha\beta\gamma\delta}(g)=0,
\end{equation}
where $C_{\alpha\beta\gamma\delta}$ is the Weyl tensor.

Eq.~\eqref{matter} suggests  that \emph{the gravitational degree of freedom should be encoded in the scalar field} $\Phi$, whereas \emph{matter should be encoded in the} $\phi_i$. We are therefore aiming at a scalar theory of gravity with actions:
\begin{eqnarray}
S_{\rm eff}=\int dx^4\sqrt{-\eta}\, \lll_{\rm eff} =S_{\rm grav}+S_{\rm matter},\label{tot-action}\\
S_{\rm grav}= \ell^2 \int dx^4 \sqrt{-\eta}\, \eta^{\mn} \partial_\mu\Phi\partial_\nu\Phi ,\label{grav-action}\\
S_{\rm matter}=\int dx^4 \sqrt{-\eta}\,\left( \sum_i\Phi^2 \eta^{\mn}\partial_\mu\phi_i\partial_\nu\phi_i + \lambda(|\phi|^2-\ell^2)\right),\label{matt-action}
\end{eqnarray}
where we have explicitly written the volume element $\sqrt{-\eta}=1$ so to make clear that these actions are given in flat spacetime.

It is easy to see that the very same actions can be recast in the form of actions in a curved spacetime endowed with the metric \eqref{conformal-metric 1}. In particular for the matter action in \eqref{matt-action} one has
\begin{equation*}
S_{\rm matter}= \int dx^4 \left( \sum_i \Phi^2 \eta^{\mn} \partial_\mu\phi_i\partial_\nu\phi_i + \lambda(|\phi|^2-\ell^2)\right) =
\end{equation*}
\begin{equation} \int \sqrt{-g}dx^4\left(\sum_ig^{\mn}\partial_\mu\phi_i\partial_\nu\phi_i + \lambda'(|\phi|^2-\ell^2)\right),\label{new Lagrangian for matter}
\end{equation}
where we  have suitably rescaled the Lagrange multiplier to $\lambda'$. This allows to construct the stress-energy tensor $T_\mn$ for the non-linear sigma model, and its trace $\textbf{T}$ with respect to the metric $g$:
\begin{equation}
T_\mn=\frac{2}{\sqrt{-g}}\frac{\delta S_{\rm matter}}{\delta g^\mn}= \sum_i \left (\partial_\mu\phi_i\partial_\nu\phi_i - \demi g_\mn (g^{\alpha\beta}\partial_\alpha\phi_i\partial_\beta\phi_i) \right),\end{equation}
\begin{equation} \textbf{T}= g^\mn T_\mn = - \Phi^{-2} \sum_i \eta^{\mn}\partial_\mu\phi_i\partial_\nu\phi_i. \nn
\end{equation}
Finally, the above result, together with the recognition that the Ricci scalar $\textbf{R}$, associated to the metric $g_{\mn}$, can be written as $\textbf{R}=  - 6 {\Box_\eta \Phi}/{\Phi^{3}}$, allows us to rewrite Eq.~\eqref{gravity EOM} as the Einstein--Fokker equation
\begin{equation}\label{einstein fokker}
\Box_\eta \Phi = \frac{1}{\ell^2}\eta^{\mn}\Phi \sum_i\partial_\mu\phi_i\partial_\nu\phi_i \quad \Leftrightarrow\quad  \textbf{R } = \frac{6}{\ell^2}  \textbf{T}.
\end{equation}

In summary, we can gather together the  equations of motion \eqref{matter}, \eqref{weyl tensor},   \eqref{einstein fokker}, obtained by introducing the metric \eqref{conformal-metric 1}, we have
\begin{eqnarray}
&& \textbf{R } =   \frac{6}{\ell^2}  \textbf{T}, \qquad C_{\alpha\beta\gamma\delta}=0. \label{emergent1}\\
&& \Box_g \phi_i + \frac{1}{\ell^{2}} g^{\mu\nu} \partial_{\mu}\phi_{j}\partial_{\nu}\phi_{k}\delta^{jk} \phi_{i}=0, \qquad |\phi|^2-\ell^2=0. \label{emergent3}
\end{eqnarray}
We recognize the equations of motion as those for Nordstr\"om gravity
\begin{equation}\label{Nord}
\textbf{R}= 24  \pi G_{\rm N}\, \textbf{T}, \qquad C_{\alpha\beta\gamma\delta}=0,
\end{equation}
coupled to a non-linear sigma model. Indeed, the rewriting of \eqref{gravity EOM}-\eqref{constraint} into the form \eqref{emergent1}-\eqref{emergent3}, is a special case of the procedure suggested by Einstein and Fokker so to cast Nordstr\"om gravity in a geometrical form \cite{fokker}.

We see from the above equation that the Newton constant $G_{\rm N}$ in our model has to be proportional to $\ell^{-2}$. However, in identifying the exact relation between the two quantities, some care has to be given to the fact that the stress-energy tensors appearing respectively in equation \eqref{emergent1} and equation \eqref{Nord} do not share the same dimensions. This is  due to the fact that the fields $\phi_i$ have the dimension of a length rather than the usual one of an energy. This implies that in order to really compare the expressions one has to suitably rescale our fields with a dimensional factor, $\Xi$, which in the end  would combine with $\ell$ so to produce an energy, ${\rm dim}[\ell\,\Xi]={\rm energy}$. In particular,  is easy to check that one has to assume ${4\pi} \ell^2 \Xi^2 \equiv  E^2_{\rm Planck}$ in order to recover the standard value of $G_{\rm N}$ (assuming $c$ as the observed speed of signals and $\hbar$ as the quantum of action).  As a final remark, we should stress that the scale $\ell$ is completely arbitrary within the emergent system and in principle should be derived from the physics of the ``atoms of spacetime" whose large N limit gives rise to \eqref{fundamental Lagrangian}.

Accidentally,  the above discussion also shows that, once the fields are suitably rescaled so to have the right dimensions,  the constraint appearing in Eq.\eqref{emergent3} is fixing the norm of the multiplet to be equal to the square of the Planck energy. This implies that the interaction terms in the aforementioned equation are indeed Planck-suppressed and hence negligible at low energy.
This should not be a surprise, given that in the end $\ell\Xi$ is the only energy scale present in our model. It is conceivable that more complicate frameworks, possibly endowed with many dimensional constants, will introduce a hierarchy of energy scales and hence break the degeneracy between the scale of gravity and the scale of matter interactions.

\subsection{Lessons from the toy model}
Initially, we have considered fields that live in a Euclidean space, and showed that there exists a class of Lagrangians such that the perturbations around some classical solutions $\bar{\psi}$ propagate in a Minkowski spacetime. In this case $\bar{\psi}$ is essentially picking up a preferred direction, so that we have a spontaneous symmetry breaking of the Euclidean symmetry. The apparent change of signature is free of the problems usually met in signature change frameworks since the theory is fundamentally Euclidean. Lorentz symmetry is only approximate, and in this sense it is \emph{emergent}.

The main lesson we want to emphasize here is that \emph{Lorentzian signature can emerge from a fundamental Euclidean theory} and this process can in principle be  reconstructed by observers living in the emergent system. In fact, while from the perturbations
point of view it is a priori difficult to see the fundamental Euclidean nature of the world, this could be guessed from the fact that some Lorentz symmetry breaking would appear at high energy (in our case in the form of a non-dynamical ether field).

In the second part of the discussion of this toy model, using a natural field redefinition adapted to the symmetries of the system, we have identified from the perturbations $\varphi_i$, a scalar field $\Phi$  encoding gravitational degrees of freedom and a set of scalar fields $\phi_i$ (a non-linear sigma model) encoding matter fields. In this sense, gravity and matter are both emergent at the same level. This approach is then rather different from the one of analogue models of gravity where one usually identifies the analogue of the gravitational  degrees of freedom with the ``background" fields, \ie the condensate or the solution $\psi$ of the equations of motion. Indeed, following this line of thought in looking for a theory of gravitational dynamics, we would be led to require that the fundamental field theory \eqref{fundamental Lagrangian} must be endowed with diffeomorphisms invariance from the very start  --- the symmetries of the background are identical by construction to the ones of the fundamental theory. This would imply that one would have  to obtain gravity from a theory which is already diffeomorphisms invariant and hence most probably with a form very close to some known theory of gravitation. For these reasons, we do expect that if an emergent picture is indeed appropriate for gravitation, then it should be of the sort presented here, with both matter and gravity emerging at the same level. {Of course, it is not possible to exclude that a full fledged theory of gravity could emerge,  together with the notion of manifold,  in a single step from the eventual semiclassical/large number limit of the fundamental objects. In this case, however, we would still have a very different picture from the one envisaged in analogue models of gravity.}

In particular, this allows not only for an emergent local Lorentz invariance for the perturbations dynamics but it leads as well to an emergent diffeomorphisms invariance. In fact, we saw how the equations of motion \eqref{gravity EOM} and \eqref{matter EOM}  could be  rewritten in a completely equivalent way using a conformally flat metric \eqref{conformal-metric 1}.  Most noticeably, they can be rewritten in an evidently diffeomorphisms invariant form, from the point of view of ``matter fields observers". In fact, following the standard hole argument (see \cite{sonego} for a careful discussion of the various issues related to diffeomorphism invariance), this also implies that the coordinates $x^\mu$, used to parameterized our theory, do not have any physical meaning from the point of view of the $\phi_i$ ``matter observers". They are merely parameters. In agreement with the fact that diffeomorphisms invariance is emergent in our system,  it can be noted that the cubic contribution \eqref{cubic contribution} ends up breaking it at the same level it breaks Lorentz invariance.

Furthermore, Nordstr\"om gravity is also a nice framework for discussing the subtle distinction between background independence and diffeomorphisms invariance \cite{bckgd}. We call background some geometrical degrees of freedom that are not dynamical. For example, in General Relativity the topology of the manifold and its dimension, or the signature of the metric, can be considered as (trivial) background quantities. We can therefore have some specific background structures while still having diffeomorphisms invariance. Nordstr\"om gravity is encoded in conformally flat metrics.
If one considers fields which are conformally coupled to the metric (such as the electromagnetic field), these fields only see the metric $\eta_\mn$ which is of course not dynamical. The Minkowski metric can be see then as a background structure, this is what one may call a ``prior geometry" (\eg see \cite{MTW}).
One may hence say that diffeomorphism invariance is somewhat of a weaker form in Nordstr\"om gravity with respect the one present in general relativity.

In particular, while the essence of diffeomorphism invariance in General Relativity is encoded in the associated Hamiltonian constraints, these are not defined in the present formulation of Nordstr\"om gravity. Furthermore,  in the most general implementations of Norstr\"om theory, quantities can be built which manifestly include the background structure $\eta_{\mn}$ and hence are not diffeomorphism invariant.
However, within our model, the prior geometry cannot be detected.
Indeed, in order to detect the Minkowski background, one should be
able to propose a method to pinpoint the conformal factor $\Phi^2$ in
the relation $g_{\mu\nu}=\Phi^2\eta_{\mu\nu}$. However, a careful
analysis shows that this is actually impossible. Let us elaborate on
this point.
If we perform a conformal transformation, $x^{\mu}\rightarrow
\bar{x}^{\mu}(x)$, the equations of motions associated to
\eqref{Lagrangian for perturbations 1} are transforming like
\begin{equation}
  \Box_{\eta}\varphi_i = 0 \rightarrow \Box_{\bar{\eta}}\varphi_i=0,
\end{equation}
where $\eta$ and $\bar{\eta}$ are two different Minkowski metrics
related by some conformal factor $\lambda(x)$. Therefore, $\eta$ and $
\bar{\eta}$ are indistinguishable, due to conformal invariance the
equations of motion for $\varphi_{i}$. Hence, what appears to be a
background structure, namely $\eta_{\mu\nu}$, is ambiguously defined,
and the coordinates $x^{\mu}$ in which the equations of motion for the
fields $\varphi_{i}$ are written have no operational meaning, they are
mere labels. Furthermore, this ambiguity in the definition of what
would be called a background structure implies an ambiguity on the
definition of the conformal factor relating the physical metric to the
would-be background structure. In this sense, within this very
specific implementation of the model which has conformal invariance,
there is no Minkowski geometry as a background. There is a background
structure, which is the conformal structure of Minkowski spacetime.
This is a mild limitation of our simple toy model as a diffeomorphism
invariant, background independent system.

Of course, the above discussion holds only at the lowest order in the
fields $\varphi_{i}$. As previously discussed, higher orders in
perturbation theory will generate terms like \eqref {cubic
contribution} producing a breaking of the conformal symmetry and hence
the appearance of the background structures, {\it i.e.} the Euclidean
space and the $\partial_{\mu}\bar{\psi}$ which have selected the
timelike direction.

Finally, Nordstr\"om gravity is only a scalar gravity theory, which has been  falsified by experiments (\eg the theory does not predict the bending of light).   In order to obtain a more physical theory, in particular General Relativity, one should surely look for more complicated emergent Lagrangians than \eqref{fundamental Lagrangian}. Of course, one would in this case aim to obtain the emergence of a theory characterized by spin-2 gravitons (while in Nordstr\"om theory the graviton is just a scalar). This would open a door to a possible conflict with the Weinberg--Witten theorem (see section \ref{sec:ww} and \cite{WW}). However, there are many ways in which such a theorem can be evaded and in particular one may guess that analogue models inspired mechanisms like the one discussed here will generically lead to Lagrangian which show Lorentz and diffeomorphism invariance only as approximate symmetries for the lowest order in the perturbative expansion.

It is unclear which sort of generalization may still lead to some viable gravitational theory from the perturbations dynamics. For example, the simple addition of a potential will in general prevent the selection of a preferred direction, except in regions where the potential is almost flat. Moreover, it would also spoil the metric interpretation of the theory. For example, the terms $|\varphi|^n$ for $n\geq1$ and $\neq 4$ cannot be rewritten as an interaction between the matter field fields $\phi$ living on the conformal metric $\Phi^2\eta_\mn$, when using the change of variables \eqref{field redefinition} (although it is interesting to note that  a $|\varphi|^4$ term would give  Nordstr\"om gravity with a cosmological constant).

Perhaps, this toy model is too simple: in order to obtain more realistic theories one has to use more ingredients. Nevertheless, the reader should be convinced, by now, that such an objective could be not completely hopeless, as this primitive toy model suggests. Of course, much more should be done in this direction.

\subsection{The role of symmetries}
It is interesting to discuss in details the features that allowed the construction of such a toy model. In particular, it is important to stress the role of symmetries, in order to make clear the way in which they enter at the various levels. As in the case of selecting Riemannian geometry out of Finsler geometry, here there are some symmetries which are absolutely essential: it is only due to their presence that we do have an emergent gravitational system possessing a geometrical nature.

We have seen that in order to produce a working model, a number of properties must be assumed. First of all, there is an underlying $ISO(4)$ symmetry
which allows us to use particularly simple affine solutions. This $ISO(4)$, when spontaneously broken, can lead to an approximate Poincar\'e invariance. Moreover, the masslessness of the resulting modes is promoting this Poincar\'e invariance to a full conformal invariance, which is approximate as well. This conformal invariance is the key symmetry which hides the background structure, forbidding a low energy observer to detect a background metric structure (there is only a background conformal structure).

Conformal invariance seems to be deeply intertwined with the possibility of writing down the resulting equations of motion in the form of a system of diffeomorphism invariant equations, as we have seen. However, in order for the Lagrangian \eqref{Lagrangian for perturbations} to be conformal invariant, there must be an overall $O(N)$ symmetry between the fields. This symmetry is just the other side of the coin of the mechanism leading to the monometricity. If two fields move in different metrics, clearly this $O(N)$ is broken and the entire model fails to provide a geometric picture, let alone a diffeo-invariant one.

In general, one should expect that in any situation in which the metric is an emergent structure, there should be a mechanism taking care of the fact that different matter fields propagate over the same geometry. In this picture, where a manifold is given from the beginning, the role of internal and spacetime symmetries is crucial. The behavior we have described is not general at all. Of course, one could conclude that this kind of models is somehow contrived and unnatural.

However, there is also a positive side: given that symmetries (both of the equations of motion and of the ground state) play a crucial
role in the emergence mechanism, the fact that our universe seems to be ruled, at large scales, by general relativity and locally by special relativity, suggests that not all the pre-geometric scenarios are viable, and that there are rather strong constraints on what are the possible mechanism of emergence.

\section{Conclusions}

In summary, we hope that the two examples of emergent gravitational dynamics presented here have suitably illustrated the potentialities of emergent gravity models inspired by the analogue gravity perspective. The first case, the one of a BEC system, has shown us that a gravitational-like dynamics (with a small cosmological constant) seems to be a natural by product of a condensation mechanism. However, the analogy with the real world was not only limited by the Lorentz breaking scale but also from the fact that gravity and matter seem to be living at rather different levels (gravity is the condensate while matter is associated to the quasi-particle states).  The toy model discussed in the second part of this proceedings is aimed at overcoming this problems. It does have Lorentz violation but only in the limit of large fluctuations of the fundamental fields $\phi_i$. This implies that the gravitational dynamics is no more endowed with a massive graviton whose mass scale is set by the UV Lorentz breaking scale of the system. Furthermore, this toy models shows how time and diffeomorphism invariance might emerge. In particular the latter is allowed by the special symmetries of the system. As we have said previously, similar symmetries are probably needed anyway to protect the IR limit of the theory from large violations of Lorentz invariance of the equations. Given that the presence of such violations seems to be a very natural way around to the Weinberg-Witten theorem obstruction, it might be that the next step towards a satisfactory emergent gravity scenario might have to consist in finding which sort of mechanism, possibly a custodial symmetry, could simultaneously guarantee the background independence of the emerging dynamics as well as a very accurate Local Lorentz invariance of the emergent spacetime. We hope to address these questions in future work.

\bibliographystyle{amsplain}

\end{document}